\documentclass[10pt,aps,prd,preprintnumbers,showpacs,superscriptaddress,nofootinbib,amsmath,amssymb,floats,floatfix,showkeys,notitlepage,longbibliography]{revtex4-2}
\usepackage{xcolor}
\usepackage{microtype}
\usepackage{hyperref}
\usepackage{orcidlink}
\hypersetup{
  colorlinks=true,
  linkcolor=blue,
  urlcolor=blue,
  citecolor=blue
}

\numberwithin{equation}{section}
\allowdisplaybreaks

\begin{document}
\sloppy
\title{Exact Spectrum of a Curvature-Adapted
\texorpdfstring{$\mathbb{Z}_{2}^{3}$}{Z2 cubed}
Dunkl--Deng--Fan/Eckart System}

\author{Nikko John Leo S. Lobos \orcidlink{0000-0001-6976-8462}}
\email{nslobos@ust.edu.ph}
\affiliation{Electronics Engineering Department, University of Santo Tomas, Espa\~na Boulevard, Sampaloc, Manila 1008, Philippines}

\begin{abstract}
The objective of this study is to construct and solve exactly a curvature-adapted radial Dunkl Hamiltonian associated with the coordinate-reflection group $\mathbb{Z}_{2}^{3}$ on three-dimensional hyperbolic space, and to determine how curvature and Dunkl multiplicities control its discrete spectrum and admissible radial states. The warped-product kinetic operator is realized through its Friedrichs extension, which selects the regular boundary condition in singular radial channels. The restriction $\alpha=2\kappa$ is an exact-solvability condition, not a general relation between molecular range and spatial curvature; it reduces the centrifugal and Deng--Fan terms without approximation to a generalized Eckart problem. We obtain the finite spectrum, Jacobi-polynomial radial eigenfunctions, normalization integrals, and strict normalizability condition $\eta_{n\ell}^{2}<q_{\kappa}/2$. Dunkl multiplicities enter the radial spectrum only through $\gamma=\mu_{1}+\mu_{2}+\mu_{3}$; reflection eigenvalues instead restrict the angular degrees to $\ell=p+2k$, with total multiplicity $2\ell+1$. Along each surviving branch, the energy increases with $\ell$ and $\gamma$, while the number of bound states decreases. Quadratic-form Hellmann--Feynman identities yield exact hyperbolic-interaction expectation values, whereas adjacent-level spacings follow from the spectrum. The undeformed and correlated flat limits recover the ordinary curvature-matched Eckart and Dunkl--Kratzer spectra, respectively. Finite-difference calculations for $\gamma=0$ and $1/2$ confirm the energies, continuum thresholds, and state counts. The construction is a tuned curvature-adapted system, not the generic Euclidean Dunkl--Deng--Fan problem. Its physical significance is to isolate, within a controlled solvable model, the effects of negative curvature and reflection deformation on level ordering, binding thresholds, and the number of bound states. The resulting formulas also provide analytic benchmarks for numerical and approximate treatments of singular curvature- and reflection-deformed radial Hamiltonians.

\end{abstract}

\keywords{Dunkl operators; hyperbolic space; Deng--Fan potential; generalized Eckart potential; Friedrichs extension;
exactly solvable quantum systems; Jacobi polynomials}

\pacs{03.65.Ge, 03.65.-w, 33.15.Mt, 02.30.Gp, 33.20.Ea}

\maketitle

\section{Introduction}
\label{sec:introduction}

Dunkl operators extend ordinary differential calculus by combining derivatives with reflections associated with a finite reflection group. For a root system with assigned multiplicities, the resulting differential--difference operators commute and generate a deformation of the ordinary Laplacian \cite{Dunkl:1989,Rosler:2003}. This framework modifies the natural integration measure, angular harmonics, and spectral decomposition of quantum Hamiltonians. For the coordinate-reflection group $\mathbb{Z}_{2}^{3}$, the three Cartesian reflections act independently, and the eigenfunctions can be classified by three reflection eigenvalues. The associated weighted polynomial spaces, spherical harmonics, and Dunkl--Fischer decomposition are developed systematically in Ref.~\cite{DunklXu:2014}.

Dunkl operators entered quantum mechanics through reflection-deformed oscillator, Coulomb, and superintegrable systems. The planar Dunkl oscillator provides an early example in which reflection symmetries modify both the separated wavefunctions and the symmetry algebra \cite{Genest:2013}. Three-dimensional Dunkl--Schr\"{o}dinger systems with central
potentials have also been separated in spherical coordinates, with
their angular functions and radial spectra expressed through classical
special functions \cite{Mota:2022DunklSpherical}. Curved extensions have since been constructed using different geometric prescriptions. Factorization methods have been applied to Dunkl oscillators on two-dimensional constant-curvature spaces \cite{Najafizade:2022}, while an exactly solvable Dunkl--Darboux III oscillator combines reflection deformation with a nonconstant-curvature geometry in arbitrary dimension \cite{Ballesteros:2024}. These models show that curvature can be incorporated consistently into selected Dunkl Hamiltonians, but the resulting kinetic operator depends on the adopted geometric construction.

A separate line of research concerns exponential molecular interactions. The Deng--Fan potential was introduced as an effective interaction for diatomic molecules \cite{DengFan:1957}. It has a finite dissociation threshold, a short-distance repulsive core, and a minimum fixed by the equilibrium separation. For nonzero angular momentum in Euclidean space, however, the exponential coordinate adapted to the potential does not rationalize the centrifugal term $r^{-2}$. Analytical treatments therefore commonly replace the centrifugal barrier by Pekeris-type, Greene--Aldrich-type, or related approximations. This procedure has been used in shifted Deng--Fan calculations of rotational--vibrational energies \cite{Hamzavi:2013,Njoku:2022}. A similar obstruction occurs in reflection-deformed systems: the Dunkl--Schr\"odinger equation for the hyperbolic P\"oschl--Teller potential has been treated using a series approximation for the inverse-square term \cite{SchulzeHalberg:2024}.

The present work follows a different construction. Rather than approximate the Euclidean centrifugal term, we define a curvature-adapted radial geometry whose angular barrier is proportional to $\operatorname{csch}^{2}(\kappa r)$. Under the restricted matching condition $\alpha=2\kappa$, this barrier and the Deng--Fan interaction become rational functions of the same exponential coordinate. The resulting system is therefore not a correction to the approximate Euclidean Dunkl--Deng--Fan spectra. It is a different Hamiltonian whose geometry and interaction are matched within a tuned exactly solvable parameter family.

We construct a three-dimensional curvature-adapted radial Dunkl model
associated with $\mathbb{Z}_{2}^{3}$. We introduce
the hyperbolic warp factor
$S_{\kappa}(r)=\sinh(\kappa r)/\kappa$, where
$\kappa>0$ is the inverse curvature radius, and
define a symmetric radial Dunkl operator in the weighted Hilbert space generated by $S_{\kappa}^{2\gamma+2}(r)$, with
$\gamma=\mu_{1}+\mu_{2}+\mu_{3}$. The operator is chosen to satisfy three conditions: it reduces to the Euclidean Dunkl Laplacian as $\kappa\rightarrow0$, becomes the ordinary hyperbolic Laplace--Beltrami operator when the multiplicities vanish, and is symmetric with respect to the warped Dunkl measure. This warped-product construction defines the model studied here; it is not claimed to be the unique covariant Dunkl operator on hyperbolic space.

The separated radial equation contains the curvature-adapted barrier
$\kappa^{2}\operatorname{csch}^{2}(\kappa r)$. Exact compatibility with the Deng--Fan interaction is obtained by imposing
$\alpha=2\kappa$. This is an analytical restriction defining the solvable subfamily, not a physical law relating molecular range to spatial curvature. Under this condition, the interaction decomposes exactly into constant, $\coth(\kappa r)$, and
$\operatorname{csch}^{2}(\kappa r)$ terms, so the complete radial equation reduces to a generalized Eckart problem \cite{Eckart:1930} without replacing the centrifugal barrier by a local approximation. The spectral solution of the generalized Eckart equation is known. The new element is the curvature-adapted Dunkl construction that produces this equation exactly from the matched geometry and interaction.

Because the combined radial interaction is singular at the origin, the operator domain forms part of the spectral problem. We define the Hamiltonian through its Friedrichs realization and classify the origin using the complete inverse-square coefficient, including both the Dunkl centrifugal contribution and the singular Deng--Fan core. This prescription selects the less singular branch in limit-circle channels and agrees with the unique admissible behavior in limit-point channels. Self-adjoint realizations of inverse-square Hamiltonians and the relevant endpoint classifications are discussed in Refs.~\cite{DerezinskiRichard:2017,GesztesyPangStanfill:2024}.

The main contributions are as follows. First, we define the curvature-adapted $\mathbb{Z}_{2}^{3}$ Dunkl operator, its weighted Hilbert space, and its Friedrichs realization. Second, we show that the matching condition $\alpha=2\kappa$ converts the full radial equation exactly into generalized Eckart form. Third, we derive the finite bound-state spectrum, Jacobi-polynomial radial eigenfunctions, normalization integrals, and strict normalizability condition. Fourth, we establish the reflection-sector selection rule
$\ell=p+2k$, derive the fixed-sector dimensions, and show that the combined angular multiplicity remains $2\ell+1$. Finally, we obtain spectral-ordering results, exact state-resolved expectation values, restricted bound-state partition sums, and an independent finite-difference check of the analytical solution.

The radial spectrum depends on the three Dunkl multiplicities only through their sum $\gamma$. Their individual values remain relevant to the weighted angular measure, generalized harmonics, and reflection-sector content. The reflection eigenvalues do not generate an independent radial interaction; instead, they determine which angular degrees occur. Along each branch that remains below the continuum threshold, the energy increases with both the angular degree $\ell$ and the total multiplicity $\gamma$, while the number of admissible radial states decreases as either parameter grows.

The exact spectrum also gives algebraic expressions for
$\langle\coth(\kappa r)\rangle$,
$\langle\operatorname{csch}^{2}(\kappa r)\rangle$, the mean matched interaction energy, the reduced radial kinetic energy, and the spacing between adjacent radial levels. Finite partition sums are defined for fixed angular channels, fixed reflection sectors, and the complete discrete bound-state space. These sums describe restricted bound-state ensembles. They exclude scattering and dissociated states and are not the full canonical thermodynamics of the unconfined system.

Two limits provide consistency checks. When the Dunkl multiplicities vanish, the construction reduces to the ordinary curvature-matched Deng--Fan/Eckart system. In the correlated flat limit
$\kappa\rightarrow0$ with $\alpha=2\kappa\rightarrow0$, the interaction becomes the Kratzer potential and the energy spectrum reduces to its Dunkl-deformed form. Independent finite-difference calculations for an undeformed channel
with $\gamma=0$ and a nonzero-Dunkl channel with $\gamma=1/2$
verify the exact energies, continuum thresholds, finite-state counts,
and the predicted change of the radial spectrum with the total Dunkl
multiplicity. For the convergence benchmark, the errors display the
second-order decrease expected from the central-difference scheme.

The paper is organized as follows. Section~\ref{sec:dunkl_kinematics} defines the curvature-adapted $\mathbb{Z}_{2}^{3}$ Dunkl operator, its weighted Hilbert space, Friedrichs realization, and reflection-sector reduction. Section~\ref{sec:curvature_matched_deng_fan} introduces the matched Deng--Fan interaction and derives the exact generalized Eckart equation. Section~\ref{sec:exact_spectrum_wavefunctions} obtains the discrete spectrum, Jacobi-polynomial eigenfunctions, normalization integrals, and finite-state condition. Section~\ref{sec:reflection_spectral_ordering} analyzes reflection-sector degeneracies, angular multiplicities, spectral ordering, and the undeformed and correlated flat limits. Section~\ref{sec:state_resolved_observables} derives exact expectation values, radial level spacings, and restricted bound-state partition sums. Section~\ref{sec:numerical_verification} provides the finite-difference verification. Section~\ref{sec:scope_limitations} states the scope and limitations of the comparison, and Section~\ref{sec:conclusion} summarizes the results.

\section{Curvature-adapted
\texorpdfstring{$\mathbb{Z}_{2}^{3}$}{Z2 cubed}
Dunkl kinematics and radial reduction}
\label{sec:dunkl_kinematics}

We consider the coordinate-reflection group
$W=\mathbb{Z}_{2}^{3}$ acting on
$\boldsymbol{x}=(x_{1},x_{2},x_{3})\in\mathbb{R}^{3}$.
Its generators reverse one Cartesian coordinate,
\[
R_{i}(x_{1},\ldots,x_{i},\ldots,x_{3})
=
(x_{1},\ldots,-x_{i},\ldots,x_{3}),
\]
and satisfy $R_{i}^{2}=1$ and $[R_{i},R_{j}]=0$.

Let
$\boldsymbol{\mu}=(\mu_{1},\mu_{2},\mu_{3})$ denote the Dunkl
multiplicities. Throughout the analysis, we assume
$\mu_{i}\geq0$. This domain gives a positive, locally integrable
Dunkl weight and the standard non-negative realization of the angular
Dunkl operator. Extending the model to
$-1/2<\mu_{i}<0$ would require a separate analysis of angular
positivity and operator domains and is not considered here. The
Cartesian Dunkl derivatives and Euclidean Dunkl Laplacian are
\begin{equation}
\mathcal{D}_{i}
=
\frac{\partial}{\partial x_{i}}
+
\frac{\mu_{i}}{x_{i}}(1-R_{i}),
\qquad
\Delta_{\boldsymbol{\mu}}
=
\sum_{i=1}^{3}\mathcal{D}_{i}^{2}.
\label{eq:dunkl_derivatives}
\end{equation}
We define the total multiplicity by
$\gamma=\mu_{1}+\mu_{2}+\mu_{3}$. We use spherical
coordinates $(r,\theta,\varphi)$, with
$\Omega=(\theta,\varphi)$, because both the central interaction and the
warp factor depend only on the geodesic radius $r$. This symmetry-adapted
choice separates the radial motion from the angular Dunkl harmonics,
diagonalizes the angular barrier in sectors of fixed $\ell$, and reduces
the spectral problem to a one-dimensional singular
Sturm--Liouville equation. In these coordinates,
the standard decomposition and angular measure are
\cite{Dunkl:1989,DunklXu:2014}
\[
\begin{aligned}
\Delta_{\boldsymbol{\mu}}
&=
\frac{\partial^{2}}{\partial r^{2}}
+
\frac{2\gamma+2}{r}\frac{\partial}{\partial r}
+
\frac{1}{r^{2}}\Delta_{S^{2},\boldsymbol{\mu}},
\\
d\omega_{\boldsymbol{\mu}}
&=
|\sin\theta\cos\varphi|^{2\mu_{1}}
|\sin\theta\sin\varphi|^{2\mu_{2}}
|\cos\theta|^{2\mu_{3}}
\sin\theta\,d\theta\,d\varphi .
\end{aligned}
\]
Here $\Delta_{\boldsymbol{\mu}}$ denotes the Euclidean
Dunkl Laplacian defined in Eq.~\eqref{eq:dunkl_derivatives}; its
subscript labels the multiplicity vector
$\boldsymbol{\mu}=(\mu_1,\mu_2,\mu_3)$. The operator
$\Delta_{S^2,\boldsymbol{\mu}}$ is its angular part on the unit sphere,
and $d\omega_{\boldsymbol{\mu}}$ is the associated weighted angular
measure. Thus $\Delta_{\boldsymbol{\mu}}$ is a single operator and not a
product or variation denoted by ``$\Delta\mu$''.

Introduce an inverse-length scale $\kappa>0$ and the warp factor
$S_{\kappa}(r)=\sinh(\kappa r)/\kappa$. When
$\boldsymbol{\mu}=\boldsymbol{0}$, this is the radial factor of the
three-dimensional hyperbolic metric
\begin{equation}
ds_{\kappa}^{2}
=
dr^{2}
+
S_{\kappa}^{2}(r)
\left(
d\theta^{2}+\sin^{2}\theta\,d\varphi^{2}
\right),
\qquad
\mathcal{R}_{\kappa}=-6\kappa^{2},
\label{eq:hyperbolic_metric}
\end{equation}
where $\kappa^{-1}$ is the curvature radius and
$\mathcal{R}_{\kappa}$ is the scalar curvature. The coordinate
reflections preserve $r$ and act as angular isometries.
In Eq.~\eqref{eq:hyperbolic_metric}, $dr^{2}$ measures
radial geodesic displacement, while
$S_{\kappa}^{2}(r)(d\theta^{2}+\sin^{2}\theta\,d\varphi^{2})$ describes
the angular two-sphere whose area grows with $r$. The constant negative
curvature $\mathcal{R}_{\kappa}=-6\kappa^{2}$ fixes the geometric length
scale and preserves the central separation used below.

Within the warped-product ansatz adopted in this work, we define the
curvature-adapted Dunkl operator by replacing the Euclidean radial
factor $r$ with $S_{\kappa}(r)$:
\begin{equation}
\begin{aligned}
\Delta_{\boldsymbol{\mu},\kappa}
&=
\frac{1}{S_{\kappa}^{2\gamma+2}(r)}
\frac{\partial}{\partial r}
\left[
S_{\kappa}^{2\gamma+2}(r)
\frac{\partial}{\partial r}
\right]
+
\frac{1}{S_{\kappa}^{2}(r)}
\Delta_{S^{2},\boldsymbol{\mu}}
\\
&=
\frac{\partial^{2}}{\partial r^{2}}
+
(2\gamma+2)\kappa\coth(\kappa r)
\frac{\partial}{\partial r}
+
\kappa^{2}\operatorname{csch}^{2}(\kappa r)
\Delta_{S^{2},\boldsymbol{\mu}}.
\end{aligned}
\label{eq:curvature_adapted_dunkl_operator}
\end{equation}
The first term in
Eq.~\eqref{eq:curvature_adapted_dunkl_operator} is the radial kinetic
operator in the warped Dunkl measure. The second term governs angular
motion and reflection-sector dependence; after angular separation it
produces the curvature-adapted centrifugal barrier proportional to
$S_{\kappa}^{-2}(r)=\kappa^{2}\operatorname{csch}^{2}(\kappa r)$.

This operator reduces to $\Delta_{\boldsymbol{\mu}}$ as
$\kappa\rightarrow0$, becomes the ordinary hyperbolic
Laplace--Beltrami operator when
$\boldsymbol{\mu}=\boldsymbol{0}$, and is symmetric with respect to
the weighted measure below. Equation
\eqref{eq:curvature_adapted_dunkl_operator} therefore defines a
specific radial deformation within the stated ansatz. It is not
claimed to be the unique covariant Dunkl Laplacian on hyperbolic
space.

The corresponding measure and Hilbert space are
\begin{equation}
d\varpi_{\boldsymbol{\mu},\kappa}
=
S_{\kappa}^{2\gamma+2}(r)\,
dr\,d\omega_{\boldsymbol{\mu}}(\Omega),
\qquad
\mathcal{H}_{\boldsymbol{\mu},\kappa}
=
L^{2}\!\left(
(0,\infty)\times S^{2},
d\varpi_{\boldsymbol{\mu},\kappa}
\right).
\label{eq:curved_dunkl_hilbert_space}
\end{equation}
Let
\[
\mathcal{C}_{\boldsymbol{\mu},\kappa}
=
C_{c}^{\infty}(0,\infty)
\otimes
\mathcal{D}_{\Omega,\boldsymbol{\mu}},
\]
where $\mathcal{D}_{\Omega,\boldsymbol{\mu}}$ is the finite linear
span of generalized Dunkl spherical harmonics. For a real, locally
integrable central potential satisfying
$V(r)\geq V_{0}$ almost everywhere, define
\begin{equation}
\widehat{H}_{\boldsymbol{\mu},\kappa}^{(0)}
=
-\frac{\hbar^{2}}{2m_{\mathrm{red}}}
\Delta_{\boldsymbol{\mu},\kappa}
+
V(r),
\qquad
m_{\mathrm{red}}>0,
\label{eq:minimal_curved_dunkl_hamiltonian}
\end{equation}
on $\mathcal{C}_{\boldsymbol{\mu},\kappa}$.
In Eq.~\eqref{eq:minimal_curved_dunkl_hamiltonian},
$-\hbar^{2}\Delta_{\boldsymbol{\mu},\kappa}/(2m_{\mathrm{red}})$ is the
curvature- and reflection-deformed kinetic energy for reduced mass
$m_{\mathrm{red}}$, whereas $V(r)$ is a central interaction depending
only on the radial coordinate. Their sum defines the minimal symmetric
Hamiltonian whose Friedrichs realization is used throughout.

Writing
$\langle\cdot,\cdot\rangle_{S^{2},\boldsymbol{\mu}}$ for the angular
inner product, integration by parts gives
\[
\begin{aligned}
h_{\boldsymbol{\mu},\kappa}^{(0)}[\Psi]
={}&
\frac{\hbar^{2}}{2m_{\mathrm{red}}}
\int_{0}^{\infty}
\left[
S_{\kappa}^{2\gamma+2}(r)
\|\partial_{r}\Psi\|_{S^{2},\boldsymbol{\mu}}^{2}
+
S_{\kappa}^{2\gamma}(r)
\left\langle
\Psi,
-\Delta_{S^{2},\boldsymbol{\mu}}\Psi
\right\rangle_{S^{2},\boldsymbol{\mu}}
\right]dr
\\
&+
\int_{(0,\infty)\times S^{2}}
V(r)|\Psi|^{2}\,
d\varpi_{\boldsymbol{\mu},\kappa}
\\
\geq{}&
V_{0}\|\Psi\|_{\boldsymbol{\mu},\kappa}^{2}.
\end{aligned}
\]
The angular operator
$-\Delta_{S^{2},\boldsymbol{\mu}}$ is non-negative for
$\mu_i\geq0$. Hence
$\widehat{H}_{\boldsymbol{\mu},\kappa}^{(0)}$ is densely defined,
symmetric, and lower semibounded. It therefore possesses a
distinguished Friedrichs extension, denoted by
$\widehat{H}_{\boldsymbol{\mu},\kappa}^{F}$. Other self-adjoint
extensions may exist in limit-circle radial channels, but the
Friedrichs realization is used throughout this work.

Because a central potential and
$\Delta_{S^{2},\boldsymbol{\mu}}$ commute with every reflection,
the eigenfunctions may be assigned reflection eigenvalues
\[
R_{i}\Psi_{\boldsymbol{\varepsilon}}
=
\varepsilon_{i}\Psi_{\boldsymbol{\varepsilon}},
\qquad
\varepsilon_{i}\in\{+1,-1\}.
\]
For a fixed sector
$\boldsymbol{\varepsilon}
=(\varepsilon_{1},\varepsilon_{2},\varepsilon_{3})$, define
$p_{i}=(1-\varepsilon_{i})/2$ and
$p=p_{1}+p_{2}+p_{3}$.

Let
$\mathcal{H}_{\ell}^{(\boldsymbol{\mu},\boldsymbol{\varepsilon})}$
be the degree-$\ell$ angular harmonic space in this sector. Its
elements satisfy
\begin{equation}
\Delta_{S^{2},\boldsymbol{\mu}}
Y_{\ell,\sigma}^{(\boldsymbol{\mu},\boldsymbol{\varepsilon})}
=
-\ell(\ell+2\gamma+1)
Y_{\ell,\sigma}^{(\boldsymbol{\mu},\boldsymbol{\varepsilon})},
\label{eq:dunkl_angular_eigenvalue}
\end{equation}
where $\sigma$ labels the fixed-sector degeneracy
\cite{DunklXu:2014}.

\noindent\textbf{Proposition II.1.}
The reflection sector
$\boldsymbol{\varepsilon}$ contains angular harmonics precisely for
\begin{equation}
\ell=p+2k,
\qquad
k=0,1,2,\ldots,
\label{eq:reflection_selection_rule}
\end{equation}
and the corresponding fixed-sector multiplicity is
\begin{equation}
\dim
\mathcal{H}_{p+2k}^{(\boldsymbol{\mu},\boldsymbol{\varepsilon})}
=
k+1.
\label{eq:reflection_sector_dimension}
\end{equation}

\noindent\textit{Proof.}
A homogeneous polynomial in the sector is odd in $x_i$ when
$p_i=1$ and even when $p_i=0$. It therefore has the form
\[
\begin{aligned}
P_{p+2k}(\boldsymbol{x})
&=
x_{1}^{p_{1}}x_{2}^{p_{2}}x_{3}^{p_{3}}
Q_{k}(x_{1}^{2},x_{2}^{2},x_{3}^{2}),
\\
\mathcal{P}_{p+2k}^{(\boldsymbol{\varepsilon})}
&=
\mathcal{H}_{p+2k}^{(\boldsymbol{\mu},\boldsymbol{\varepsilon})}
\oplus
r^{2}
\mathcal{P}_{p+2k-2}^{(\boldsymbol{\varepsilon})},
\end{aligned}
\]
where the second line is the sector-preserving Dunkl--Fischer
decomposition \cite{DunklXu:2014}. The two polynomial spaces have
dimensions $\binom{k+2}{2}$ and $\binom{k+1}{2}$, respectively, with
the second space absent when $k=0$. Their difference is $k+1$, proving
both statements.

The reflection signs therefore restrict the allowed angular degrees
but do not generate an additional radial interaction. The minimum
degree in a sector is $\ell_{\min}=p$.

For fixed $(\ell,\boldsymbol{\varepsilon})$, write
\[
\Psi(r,\Omega)
=
F_{\ell}(r)
Y_{\ell,\sigma}^{(\boldsymbol{\mu},\boldsymbol{\varepsilon})}(\Omega)
\]
and define
$u_{\ell}(r)=S_{\kappa}^{\gamma+1}(r)F_{\ell}(r)$. Using
$S_{\kappa}'/S_{\kappa}=\kappa\coth(\kappa r)$ removes the first radial
derivative. With
$\Lambda_{\ell,\gamma}
=(\ell+\gamma)(\ell+\gamma+1)$, the reduced equation becomes
\begin{equation}
\left[
-\frac{\hbar^{2}}{2m_{\mathrm{red}}}
\frac{d^{2}}{dr^{2}}
+
\frac{\hbar^{2}\kappa^{2}}{2m_{\mathrm{red}}}
(\gamma+1)^{2}
+
\frac{\hbar^{2}\kappa^{2}}{2m_{\mathrm{red}}}
\Lambda_{\ell,\gamma}
\operatorname{csch}^{2}(\kappa r)
+
V(r)
\right]
u_{\ell}(r)
=
Eu_{\ell}(r).
\label{eq:reduced_radial_equation}
\end{equation}
The four terms on the left-hand side of
Eq.~\eqref{eq:reduced_radial_equation} have distinct roles: the second
derivative is the one-dimensional radial kinetic energy,
$\hbar^{2}\kappa^{2}(\gamma+1)^{2}/(2m_{\mathrm{red}})$ is a constant
geometric shift,
$\hbar^{2}\kappa^{2}\Lambda_{\ell,\gamma}
\operatorname{csch}^{2}(\kappa r)/(2m_{\mathrm{red}})$ is the
curvature-adapted centrifugal term, and $V(r)$ is the central
interaction. The eigenvalue $E$ is the total energy.

The transformation is unitary at the radial level:
\[
\int_{0}^{\infty}
|F_{\ell}(r)|^{2}
S_{\kappa}^{2\gamma+2}(r)\,dr
=
\int_{0}^{\infty}
|u_{\ell}(r)|^{2}\,dr.
\]
At fixed $\ell$, the radial equation depends on the three
multiplicities only through $\gamma$. Their individual values remain
relevant to the angular measure, angular harmonics, reflection-sector
content, and degeneracies.

The endpoint classification at $r=0$ must include any inverse-square
singularity contributed by the central potential. Suppose
\begin{equation}
V(r)
=
\frac{\hbar^{2}}{2m_{\mathrm{red}}}
\frac{g_{V}}{r^{2}}
+
o(r^{-2}),
\qquad
r\rightarrow0^{+},
\label{eq:potential_inverse_square_coefficient}
\end{equation}
where $g_V$ is the dimensionless coefficient of the inverse-square
singularity contributed by the central potential. The geometric factor
has the short-distance expansion
\begin{equation}
\kappa^{2}\operatorname{csch}^{2}(\kappa r)
=
\frac{1}{r^{2}}
-
\frac{\kappa^{2}}{3}
+
O(\kappa^{4}r^{2}),
\qquad r\rightarrow0^{+}.
\label{eq:csch_short_distance_expansion}
\end{equation}
The notation $O(1)$ in the abbreviated relation
$\kappa^{2}\operatorname{csch}^{2}(\kappa r)=r^{-2}+O(1)$ denotes the
part that remains bounded as $r\to0$; its leading value is
$-\kappa^{2}/3$. It is a regular geometric remainder and is not related
to $g_V$. The coefficient $g_V$ multiplies the separate $r^{-2}$
singularity of $V(r)$. Consequently, collecting only the two
inverse-square terms gives
$[\Lambda_{\ell,\gamma}+g_V]/r^{2}$.

The complete local coupling is therefore
\begin{equation}
g_{\ell,\gamma}
=
\Lambda_{\ell,\gamma}+g_{V}.
\label{eq:full_inverse_square_coupling}
\end{equation}
For $g_{\ell,\gamma}>-1/4$, the two local powers are
$r^{1/2\pm\nu_{\ell,\gamma}}$, where
$\nu_{\ell,\gamma}
=\sqrt{1/4+g_{\ell,\gamma}}$. The origin is limit circle for
$-1/4\leq g_{\ell,\gamma}<3/4$ and limit point for
$g_{\ell,\gamma}\geq3/4$. At the critical coupling
$g_{\ell,\gamma}=-1/4$, the second independent solution is
proportional to $r^{1/2}\ln r$. The Friedrichs extension selects
\begin{equation}
u_{\ell}(r)
=
O\!\left(
r^{1/2+\nu_{\ell,\gamma}}
\right),
\qquad
r\rightarrow0^{+},
\label{eq:friedrichs_radial_behavior}
\end{equation}
and excludes the logarithmic branch
\cite{DerezinskiRichard:2017,GesztesyPangStanfill:2024}. Values
$g_{\ell,\gamma}<-1/4$ are excluded because the inverse-square
quadratic form is not lower semibounded.

For
$-1/4<g_{\ell,\gamma}<3/4$, both local powers are square-integrable,
but the Friedrichs domain sets the coefficient of
$r^{1/2-\nu_{\ell,\gamma}}$ to zero and retains
$r^{1/2+\nu_{\ell,\gamma}}$. At
$g_{\ell,\gamma}=-1/4$, it retains the $r^{1/2}$ solution and sets the
coefficient of $r^{1/2}\ln r$ to zero. For
$g_{\ell,\gamma}\geq3/4$, the minus branch is not square-integrable,
so the plus branch is the unique admissible endpoint behavior.

The curvature-dependent terms in
Eq.~\eqref{eq:reduced_radial_equation} are the geometric shift
\[
E_{\mathrm{geom}}
=
\frac{\hbar^{2}\kappa^{2}}{2m_{\mathrm{red}}}
(\gamma+1)^{2}
\]
and the centrifugal contribution
\[
V_{\ell,\gamma}^{\mathrm{cent}}(r)
=
\frac{\hbar^{2}\kappa^{2}}{2m_{\mathrm{red}}}
\Lambda_{\ell,\gamma}
\operatorname{csch}^{2}(\kappa r).
\]
Near the origin,
\[
\kappa^{2}\operatorname{csch}^{2}(\kappa r)
=
\frac{1}{r^{2}}
-
\frac{\kappa^{2}}{3}
+
O(\kappa^{4}r^{2}),
\]
whereas at large radius,
\[
\kappa^{2}\operatorname{csch}^{2}(\kappa r)
=
4\kappa^{2}e^{-2\kappa r}
+
O(e^{-4\kappa r}).
\]
The curved centrifugal barrier therefore retains the Euclidean
inverse-square singularity near the origin but decays exponentially at
large radius.

For fixed $r$, the flat-curvature limit is
\begin{equation}
\lim_{\kappa\rightarrow0}
\Delta_{\boldsymbol{\mu},\kappa}
=
\Delta_{\boldsymbol{\mu}},
\qquad
\lim_{\kappa\rightarrow0}
\left[
E_{\mathrm{geom}}
+
V_{\ell,\gamma}^{\mathrm{cent}}(r)
\right]
=
\frac{\hbar^{2}}{2m_{\mathrm{red}}}
\frac{\Lambda_{\ell,\gamma}}{r^{2}}.
\label{eq:flat_operator_limit}
\end{equation}
Accordingly, Eq.~\eqref{eq:reduced_radial_equation} reduces to
\[
\left[
-\frac{\hbar^{2}}{2m_{\mathrm{red}}}
\frac{d^{2}}{dr^{2}}
+
\frac{\hbar^{2}}{2m_{\mathrm{red}}}
\frac{\Lambda_{\ell,\gamma}}{r^{2}}
+
V(r)
\right]
u_{\ell}(r)
=
Eu_{\ell}(r),
\]
which is the Euclidean radial Dunkl equation. In the undeformed limit
$\mu_i\rightarrow0$, one has
$\gamma\rightarrow0$ and
$\Lambda_{\ell,\gamma}\rightarrow\ell(\ell+1)$. Equation
\eqref{eq:reduced_radial_equation} then becomes the ordinary reduced
radial Schr\"odinger equation on the hyperbolic metric
\eqref{eq:hyperbolic_metric}.

The central result of this section is
Eq.~\eqref{eq:reduced_radial_equation}. Its angular barrier reproduces
the Euclidean inverse-square behavior near the origin and develops an
exponentially decaying hyperbolic tail. The radial dynamics depend on
the multiplicities through their sum $\gamma$, while the individual
$\mu_i$ remain encoded in the angular measure, harmonics, reflection
sectors, and degeneracies.

\section{Curvature-matched Deng--Fan interaction and exact Eckart reduction}
\label{sec:curvature_matched_deng_fan}
The reduced Hamiltonian in
Eq.~\eqref{eq:reduced_radial_equation} contains the curvature-adapted
centrifugal factor
$\kappa^{2}\operatorname{csch}^{2}(\kappa r)$. It has the Euclidean
inverse-square behavior near the origin but decays exponentially at
large radius. This functional form can be combined exactly with a
matched Deng--Fan interaction, thereby avoiding the local centrifugal
approximations used in Euclidean rotational treatments
\cite{Hamzavi:2013,Njoku:2022}. The resulting model is a tuned
curvature-adapted Dunkl Hamiltonian, not the generic Euclidean
Dunkl--Deng--Fan system.

The Deng--Fan potential is
\[
V_{\mathrm{DF}}(r)
=
D_{e}
\left[
1-\frac{b}{e^{\alpha r}-1}
\right]^{2},
\qquad
b=e^{\alpha r_{e}}-1,
\]
where $D_{e}>0$ is the dissociation energy, $r_{e}>0$ is the
equilibrium separation, and $\alpha>0$ is the inverse range
\cite{DengFan:1957}. This parametrization gives
$V_{\mathrm{DF}}(r_{e})=0$ and
$V_{\mathrm{DF}}(r)\rightarrow D_{e}$ as
$r\rightarrow\infty$.

For independent $\alpha$ and $\kappa$, the interaction and the
curvature-adapted centrifugal term are not rational functions of the
same exponential coordinate. We therefore restrict the model to
\begin{equation}
\alpha=2\kappa,
\qquad
\kappa>0.
\label{eq:curvature_range_matching_condition}
\end{equation}
Equation~\eqref{eq:curvature_range_matching_condition} is an
exact-solvability condition defining the model considered below. It is
not a molecular law or a general physical relation between interaction
range and spatial curvature. Once this condition is imposed,
$\alpha$ is no longer independent and
$b=e^{2\kappa r_{e}}-1>0$. The matched interaction is
\begin{equation}
V_{\mathrm{DF}}^{(\kappa)}(r)
=
D_{e}
\left[
1-\frac{b}{e^{2\kappa r}-1}
\right]^{2}.
\label{eq:curvature_matched_deng_fan_potential}
\end{equation}
The potential is non-negative because it is the square of a real
function multiplied by $D_{e}>0$. Its derivative is
\[
\frac{dV_{\mathrm{DF}}^{(\kappa)}}{dr}
=
\frac{4\kappa D_{e}b e^{2\kappa r}}
{(e^{2\kappa r}-1)^{2}}
\left[
1-\frac{b}{e^{2\kappa r}-1}
\right].
\]
The prefactor is positive for $r>0$, while the bracket is negative for
$r<r_{e}$, vanishes at $r=r_{e}$, and is positive for $r>r_{e}$.
Therefore, $r_{e}$ is the unique global minimum,
$V_{\mathrm{DF}}^{(\kappa)}(r_{e})=0$, and
$V_{\mathrm{DF}}^{(\kappa)}(r)\rightarrow D_{e}$ as
$r\rightarrow\infty$.

Near the origin,
\[
V_{\mathrm{DF}}^{(\kappa)}(r)
=
\frac{D_{e}b^{2}}{4\kappa^{2}r^{2}}
-
\frac{D_{e}b(b+2)}{2\kappa r}
+
O(1).
\]
The leading term contributes
\[
g_{V}
=
\frac{m_{\mathrm{red}}D_{e}b^{2}}
{2\hbar^{2}\kappa^{2}}
\]
to the dimensionless inverse-square coefficient defined in
Eq.~\eqref{eq:potential_inverse_square_coefficient}. The attractive
$r^{-1}$ term is subleading and does not modify the endpoint
classification.

The identities
\[
\frac{1}{e^{2\kappa r}-1}
=
\frac{1}{2}
\left[
\coth(\kappa r)-1
\right],
\qquad
\coth^{2}(\kappa r)
=
1+\operatorname{csch}^{2}(\kappa r)
\]
give the exact decomposition
\begin{equation}
\begin{aligned}
V_{\mathrm{DF}}^{(\kappa)}(r)
={}&
D_{e}
\left(
1+b+\frac{b^{2}}{2}
\right)
-
\frac{D_{e}b(b+2)}{2}
\coth(\kappa r)
\\
&+
\frac{D_{e}b^{2}}{4}
\operatorname{csch}^{2}(\kappa r).
\end{aligned}
\label{eq:deng_fan_exact_eckart_decomposition}
\end{equation}
This is an exact algebraic identity. No replacement of $r^{-2}$,
Pekeris-type approximation, or asymptotic fitting is used.

Equation~\eqref{eq:deng_fan_exact_eckart_decomposition} has the
functional form of a generalized Eckart interaction
\cite{Eckart:1930,Quesne:2012Eckart}. The generalized Eckart equation,
including its finite bound-state spectrum and Jacobi-polynomial
solutions, is a known exactly solvable problem
\cite{Quesne:2012Eckart}. The new element here is the
curvature-adapted Dunkl construction that produces this equation
exactly from the matched Deng--Fan interaction and the hyperbolic
centrifugal term. The contribution is therefore the construction and
spectral realization of the tuned Hamiltonian, not the discovery of a
new solution method for the standard Eckart system.

The common solvable structure is explicit in the coordinate
$z=e^{-2\kappa r}\in(0,1)$:
\[
\operatorname{csch}^{2}(\kappa r)
=
\frac{4z}{(1-z)^{2}},
\qquad
\coth(\kappa r)
=
\frac{1+z}{1-z},
\qquad
V_{\mathrm{DF}}^{(\kappa)}(z)
=
D_{e}
\frac{[1-(1+b)z]^{2}}{(1-z)^{2}}.
\]
Thus, both the curvature-adapted centrifugal term and the interaction
are rational functions of the same variable. This identifies one
directly solvable matched subfamily but does not establish that
Eq.~\eqref{eq:curvature_range_matching_condition} is the only possible
solvability condition.

Substituting
Eq.~\eqref{eq:deng_fan_exact_eckart_decomposition} into
Eq.~\eqref{eq:reduced_radial_equation}, define
\begin{equation}
\mathcal{A}_{\ell,\gamma}
=
\frac{\hbar^{2}\kappa^{2}}{2m_{\mathrm{red}}}
\Lambda_{\ell,\gamma}
+
\frac{D_{e}b^{2}}{4},
\qquad
\mathcal{B}_{\kappa}
=
\frac{D_{e}b(b+2)}{2},
\label{eq:eckart_csch_coth_coefficients}
\end{equation}
and
\begin{equation}
\mathcal{E}_{\mathrm{c}}
=
\frac{\hbar^{2}\kappa^{2}}{2m_{\mathrm{red}}}
(\gamma+1)^{2}
+
D_{e}
\left(
1+b+\frac{b^{2}}{2}
\right).
\label{eq:constant_energy_shift}
\end{equation}
The reduced radial equation becomes
\begin{equation}
\left[
-\frac{\hbar^{2}}{2m_{\mathrm{red}}}
\frac{d^{2}}{dr^{2}}
+
\mathcal{A}_{\ell,\gamma}
\operatorname{csch}^{2}(\kappa r)
-
\mathcal{B}_{\kappa}
\coth(\kappa r)
\right]
u_{\ell}(r)
=
\left(
E-\mathcal{E}_{\mathrm{c}}
\right)
u_{\ell}(r).
\label{eq:exact_eckart_radial_equation}
\end{equation}
The individual multiplicities enter the radial problem only through
$\gamma=\mu_{1}+\mu_{2}+\mu_{3}$ and
$\Lambda_{\ell,\gamma}=(\ell+\gamma)(\ell+\gamma+1)$. Their separate
values remain relevant to the weighted angular measure, angular
harmonics, reflection sectors, and degeneracies.

The dimensionless couplings are
\begin{equation}
\begin{aligned}
a_{\ell,\gamma}
&=
\frac{2m_{\mathrm{red}}\mathcal{A}_{\ell,\gamma}}
{\hbar^{2}\kappa^{2}}
=
\Lambda_{\ell,\gamma}
+
\frac{m_{\mathrm{red}}D_{e}b^{2}}
{2\hbar^{2}\kappa^{2}},
\\
q_{\kappa}
&=
\frac{2m_{\mathrm{red}}\mathcal{B}_{\kappa}}
{\hbar^{2}\kappa^{2}}
=
\frac{m_{\mathrm{red}}D_{e}b(b+2)}
{\hbar^{2}\kappa^{2}}.
\end{aligned}
\label{eq:dimensionless_eckart_couplings}
\end{equation}

Each coupling in
Eq.~\eqref{eq:dimensionless_eckart_couplings} has been defined at its
first appearance. For convenient reference, the complete set of main
parameters, couplings, and sign/range conventions is also collected in
Appendix~\ref{app:eckart_parameter_summary}.

For
$\mu_{i}\geq0$, $\ell\in\mathbb{N}_{0}$, and
$D_{e},r_{e},\kappa>0$, one has
$a_{\ell,\gamma}>0$ and $q_{\kappa}>0$. Moreover,
$a_{\ell,\gamma}$ is the complete inverse-square coupling
$g_{\ell,\gamma}$ defined in
Eq.~\eqref{eq:full_inverse_square_coupling}; it includes both the
Dunkl centrifugal term and the singular Deng--Fan core.

The leading short-distance equation is
\[
-u_{\ell}''(r)
+
\frac{a_{\ell,\gamma}}{r^{2}}
u_{\ell}(r)
=
0.
\]
Its powers satisfy $s(s-1)=a_{\ell,\gamma}$. The Friedrichs branch is
therefore
\begin{equation}
u_{\ell}(r)
=
O\left(
r^{\lambda_{\ell,\gamma}}
\right),
\qquad
\lambda_{\ell,\gamma}
=
\frac{1}{2}
+
\sqrt{
\frac{1}{4}+a_{\ell,\gamma}
},
\qquad
r\rightarrow0^{+}.
\label{eq:regular_origin_boundary_condition}
\end{equation}
Because $a_{\ell,\gamma}>0$, the inverse-square form is lower
semibounded and no fall-to-the-center instability occurs. The origin
is limit circle for
$0<a_{\ell,\gamma}<3/4$ and limit point for
$a_{\ell,\gamma}\geq3/4$. In the limit-circle range,
Eq.~\eqref{eq:regular_origin_boundary_condition} selects the
Friedrichs extension. In the limit-point range, the admissible
behavior is fixed without an additional boundary condition.

Thus, in the matched
problem the selected endpoint domain retains
$u_\ell(r)\sim r^{\lambda_{\ell,\gamma}}$ and sets the coefficient of
the second local power
$r^{1-\lambda_{\ell,\gamma}}$ to zero. Since
$1-e^{-2\kappa r}=2\kappa r+\mathcal{O}(r^2)$, the same choice becomes
$(1-z)^{\lambda_{\ell,\gamma}}$ rather than
$(1-z)^{1-\lambda_{\ell,\gamma}}$ in the hypergeometric coordinate.

At large radius,
$\operatorname{csch}^{2}(\kappa r)=O(e^{-2\kappa r})$ and
$\coth(\kappa r)=1+O(e^{-2\kappa r})$. The effective radial potential
therefore approaches
\begin{equation}
E_{\mathrm{th}}
=
\mathcal{E}_{\mathrm{c}}
-
\mathcal{B}_{\kappa}
=
D_{e}
+
\frac{\hbar^{2}\kappa^{2}}{2m_{\mathrm{red}}}
(\gamma+1)^{2}.
\label{eq:curved_dunkl_continuum_threshold}
\end{equation}
A discrete state must satisfy
$E<E_{\mathrm{th}}$. Equality corresponds to the continuum threshold
and does not define a normalizable bound state.

The flat-curvature limit must be taken along the matched family.
Because $\alpha=2\kappa$, the limit
$\kappa\rightarrow0$ also forces $\alpha\rightarrow0$. For fixed
$r>0$,
\[
\frac{e^{2\kappa r_{e}}-1}
{e^{2\kappa r}-1}
\longrightarrow
\frac{r_{e}}{r},
\]
and hence
\[
V_{\mathrm{DF}}^{(\kappa)}(r)
\longrightarrow
D_{e}
\left(
1-\frac{r_{e}}{r}
\right)^{2}.
\]
Using Eq.~\eqref{eq:flat_operator_limit}, the reduced equation becomes
\begin{equation}
\left[
-\frac{\hbar^{2}}{2m_{\mathrm{red}}}
\frac{d^{2}}{dr^{2}}
+
\frac{\hbar^{2}}{2m_{\mathrm{red}}}
\frac{\Lambda_{\ell,\gamma}}{r^{2}}
+
D_{e}
\left(
1-\frac{r_{e}}{r}
\right)^{2}
\right]
u_{\ell}(r)
=
Eu_{\ell}(r).
\label{eq:curvature_matched_flat_limit_radial_equation}
\end{equation}
This is the Dunkl--Kratzer radial problem. It is not the Euclidean
Dunkl--Deng--Fan problem with a fixed nonzero inverse range
$\alpha$. The limit is therefore correlated and applies to the tuned
curvature-matched family.

The central result of this section is
Eq.~\eqref{eq:exact_eckart_radial_equation}. It follows exactly from
the warped-product Dunkl geometry and the matching condition
\eqref{eq:curvature_range_matching_condition}. Its endpoint behavior
is controlled by the complete positive coupling in
Eq.~\eqref{eq:dimensionless_eckart_couplings}, while its asymptotic
threshold is given by
Eq.~\eqref{eq:curved_dunkl_continuum_threshold}. The exact spectrum
and normalizable radial eigenfunctions are derived in the next
section.

\section{Exact spectrum and Jacobi-polynomial eigenfunctions}
\label{sec:exact_spectrum_wavefunctions}
We solve the generalized Eckart equation
\eqref{eq:exact_eckart_radial_equation} in the curvature-matched sector
$\alpha=2\kappa$. The quantities
$\mathcal{A}_{\ell,\gamma}$,
$\mathcal{B}_{\kappa}$,
$\mathcal{E}_{\mathrm{c}}$,
$a_{\ell,\gamma}$, and
$q_{\kappa}$ are defined in
Eqs.~\eqref{eq:eckart_csch_coth_coefficients}--
\eqref{eq:dimensionless_eckart_couplings}. The parameter domain of
Section~\ref{sec:curvature_matched_deng_fan} gives
$a_{\ell,\gamma}>0$ and $q_{\kappa}>0$.

Set
$x=\kappa r$ and
$U_{\ell}(x)=u_{\ell}(x/\kappa)$. The dimensionless energy is
\begin{equation}
\epsilon
=
\frac{2m_{\mathrm{red}}}{\hbar^{2}\kappa^{2}}
\left(
E-\mathcal{E}_{\mathrm{c}}
\right).
\label{eq:dimensionless_energy}
\end{equation}
Equation~\eqref{eq:exact_eckart_radial_equation} becomes
\begin{equation}
\left[
-\frac{d^{2}}{dx^{2}}
+
a_{\ell,\gamma}\operatorname{csch}^{2}x
-
q_{\kappa}\coth x
\right]
U_{\ell}(x)
=
\epsilon U_{\ell}(x).
\label{eq:dimensionless_eckart_equation}
\end{equation}
Introduce
$z=e^{-2x}=e^{-2\kappa r}$, with $0<z<1$. The endpoints
$z=0$ and $z=1$ correspond to
$r\rightarrow\infty$ and $r\rightarrow0^{+}$, respectively. Using
\[
\frac{d}{dx}
=
-2z\frac{d}{dz},
\qquad
\operatorname{csch}^{2}x
=
\frac{4z}{(1-z)^{2}},
\qquad
\coth x
=
\frac{1+z}{1-z},
\]
Eq.~\eqref{eq:dimensionless_eckart_equation} becomes
\begin{equation}
\frac{d^{2}U_{\ell}}{dz^{2}}
+
\frac{1}{z}\frac{dU_{\ell}}{dz}
+
\frac{
\epsilon(1-z)^{2}
+
q_{\kappa}(1-z^{2})
-
4a_{\ell,\gamma}z
}{
4z^{2}(1-z)^{2}
}
U_{\ell}
=
0.
\label{eq:hypergeometric_pre_reduction_equation}
\end{equation}
At $z\rightarrow0$, the decaying solution behaves as
$U_{\ell}\sim z^{\rho}$, where
\[
\rho
=
\frac{1}{2}
\sqrt{-\epsilon-q_{\kappa}}.
\]
A bound state requires $\rho>0$, or equivalently
$\epsilon<-q_{\kappa}$ and
$E<E_{\mathrm{th}}$.

At $z\rightarrow1$, the two local powers are
$(1-z)^{\lambda_{\ell,\gamma}}$ and
$(1-z)^{1-\lambda_{\ell,\gamma}}$, where
\[
\lambda_{\ell,\gamma}
=
\frac{1}{2}
+
\sqrt{
\frac{1}{4}+a_{\ell,\gamma}
}.
\]
This is the Friedrichs exponent obtained in
Eq.~\eqref{eq:regular_origin_boundary_condition}. Since
$1-z=2\kappa r+\mathcal{O}(r^{2})$ near the origin, the first branch gives
$u_{\ell}(r)\sim r^{\lambda_{\ell,\gamma}}$ and satisfies the selected
boundary condition.

Factor out both endpoint behaviors:
\[
U_{\ell}(z)
=
z^{\rho}
(1-z)^{\lambda_{\ell,\gamma}}
f_{\ell}(z).
\]
Using
$4\rho^{2}=-\epsilon-q_{\kappa}$ and
$\lambda_{\ell,\gamma}
(\lambda_{\ell,\gamma}-1)=a_{\ell,\gamma}$ gives
\begin{equation}
\begin{aligned}
z(1-z)f_{\ell}''
&+
\left[
1+2\rho
-
\left(
1+2\rho+2\lambda_{\ell,\gamma}
\right)z
\right]f_{\ell}'
\\
&-
\left[
\lambda_{\ell,\gamma}
\left(
\lambda_{\ell,\gamma}+2\rho
\right)
-
\frac{q_{\kappa}}{2}
\right]f_{\ell}
=
0.
\end{aligned}
\label{eq:gauss_hypergeometric_equation}
\end{equation}
The negative sign multiplying $q_{\kappa}/2$ inside the square
bracket follows directly from
Eq.~\eqref{eq:hypergeometric_pre_reduction_equation}. This sign is
required for consistency with the hypergeometric parameters,
quantization condition, and energy spectrum.

Equation~\eqref{eq:gauss_hypergeometric_equation} has the Gauss form
with
\[
\begin{aligned}
\mathsf{A}
&=
\rho+\lambda_{\ell,\gamma}
-
\sqrt{
\rho^{2}+\frac{q_{\kappa}}{2}
},
\\
\mathsf{B}
&=
\rho+\lambda_{\ell,\gamma}
+
\sqrt{
\rho^{2}+\frac{q_{\kappa}}{2}
},
\\
\mathsf{C}
&=
1+2\rho .
\end{aligned}
\]
The solution selected at $z=0$ is
${}_{2}F_{1}(\mathsf{A},\mathsf{B};\mathsf{C};z)$.
Near $z=1$, the full radial function contains the Friedrichs and
non-Friedrichs powers
$(1-z)^{\lambda_{\ell,\gamma}}$ and
$(1-z)^{1-\lambda_{\ell,\gamma}}$. The coefficient of the latter is
proportional to
\[
\frac{
\Gamma(\mathsf{C})
\Gamma(\mathsf{A}+\mathsf{B}-\mathsf{C})
}{
\Gamma(\mathsf{A})\Gamma(\mathsf{B})
}.
\]
Because $\mathsf{B}>0$, the non-Friedrichs coefficient vanishes when
\begin{equation}
\mathsf{A}=-n,
\qquad
n=0,1,2,\ldots .
\label{eq:hypergeometric_termination_condition}
\end{equation}
This condition also terminates the hypergeometric series.

Define
\begin{equation}
\eta_{n\ell}
=
n+\lambda_{\ell,\gamma}.
\label{eq:principal_radial_parameter}
\end{equation}
The termination condition gives
\[
\sqrt{
\rho_{n\ell}^{2}
+
\frac{q_{\kappa}}{2}
}
=
\rho_{n\ell}+\eta_{n\ell}.
\]
Both sides are positive, so squaring introduces no additional branch.
Solving for the decay exponent gives
\begin{equation}
\rho_{n\ell}
=
\frac{q_{\kappa}}{4\eta_{n\ell}}
-
\frac{\eta_{n\ell}}{2}.
\label{eq:exact_decay_exponent}
\end{equation}

Equation~\eqref{eq:hypergeometric_termination_condition} has already
removed the non-Friedrichs
$(1-z)^{1-\lambda_{\ell,\gamma}}$ branch at the origin. The remaining
independent requirement for a discrete state is therefore
$\rho_{n\ell}>0$, which enforces exponential decay at infinity; the
case $\rho_{n\ell}=0$ lies at the continuum threshold.

The normalizability condition $\rho_{n\ell}>0$ is equivalent to the
strict inequality
\begin{equation}
\eta_{n\ell}^{2}
<
\frac{q_{\kappa}}{2}.
\label{eq:exact_bound_state_condition}
\end{equation}

\noindent\textbf{Proposition IV.1.}
For every pair $(n,\ell)$ satisfying
Eq.~\eqref{eq:exact_bound_state_condition}, the Friedrichs radial
Hamiltonian has the bound-state energy
\begin{equation}
E_{n\ell}
=
\mathcal{E}_{\mathrm{c}}
-
\frac{\hbar^{2}\kappa^{2}}{2m_{\mathrm{red}}}
\left[
\eta_{n\ell}^{2}
+
\frac{q_{\kappa}^{2}}{4\eta_{n\ell}^{2}}
\right].
\label{eq:exact_energy_spectrum}
\end{equation}
\noindent\textit{Proof.}
The definition of $\rho$ gives
\[
\epsilon_{n\ell}
=
-q_{\kappa}-4\rho_{n\ell}^{2}.
\]
Substitution of Eq.~\eqref{eq:exact_decay_exponent} yields
\[
\epsilon_{n\ell}
=
-\eta_{n\ell}^{2}
-
\frac{q_{\kappa}^{2}}{4\eta_{n\ell}^{2}}.
\]
Equation~\eqref{eq:dimensionless_energy} then gives
Eq.~\eqref{eq:exact_energy_spectrum}. The strict inequality
\eqref{eq:exact_bound_state_condition} ensures exponential decay at
infinity, while
Eq.~\eqref{eq:hypergeometric_termination_condition} removes the
non-Friedrichs branch at the origin. The resulting function therefore
belongs to the domain of the selected self-adjoint Hamiltonian.
\hfill$\square$

Using
\[
E_{\mathrm{th}}
=
\mathcal{E}_{\mathrm{c}}
-
\frac{\hbar^{2}\kappa^{2}}{2m_{\mathrm{red}}}
q_{\kappa},
\]
the separation from the continuum is
\begin{equation}
E_{\mathrm{th}}-E_{n\ell}
=
\frac{\hbar^{2}\kappa^{2}}{2m_{\mathrm{red}}}
\frac{
\left(
q_{\kappa}-2\eta_{n\ell}^{2}
\right)^{2}
}{
4\eta_{n\ell}^{2}
}.
\label{eq:threshold_energy_difference}
\end{equation}
This quantity is strictly positive for every state satisfying
Eq.~\eqref{eq:exact_bound_state_condition}. At
$\eta_{n\ell}^{2}=q_{\kappa}/2$, one has
$\rho_{n\ell}=0$ and
$E_{n\ell}=E_{\mathrm{th}}$. The radial function then approaches a
nonzero constant at infinity and is not square-integrable. The
equality case is therefore excluded from the discrete spectrum.

Under the termination condition, the hypergeometric function reduces
to a Jacobi polynomial. In the physical radial coordinate, the
eigenfunction is
\begin{equation}
\begin{aligned}
u_{n\ell}(r)
={}&
\mathcal{N}_{n\ell}
e^{-2\kappa\rho_{n\ell}r}
\left(
1-e^{-2\kappa r}
\right)^{\lambda_{\ell,\gamma}}
\\
&\times
P_{n}^{\left(
2\rho_{n\ell},
2\lambda_{\ell,\gamma}-1
\right)}
\left(
1-2e^{-2\kappa r}
\right).
\end{aligned}
\label{eq:exact_jacobi_radial_wavefunction_r}
\end{equation}
The termination condition also implies
\[
\lambda_{\ell,\gamma}
\left(
\lambda_{\ell,\gamma}+2\rho_{n\ell}
\right)
-
\frac{q_{\kappa}}{2}
=
-n
\left(
n+2\rho_{n\ell}+2\lambda_{\ell,\gamma}
\right).
\]
Substitution into the Jacobi differential equation reproduces
Eq.~\eqref{eq:gauss_hypergeometric_equation}. This provides a direct
analytical check of the radial eigenfunction and the sign of the
$q_{\kappa}$ term.

The reduced radial normalization convention is
\begin{equation}
\int_{0}^{\infty}
|u_{n\ell}(r)|^{2}\,dr
=
1.
\label{eq:radial_normalization_condition}
\end{equation}
With $z=e^{-2\kappa r}$, the normalization constant is
$|\mathcal{N}_{n\ell}|^{2}=2\kappa/\mathcal{I}_{n\ell}$, where
\[
\begin{aligned}
\mathcal{I}_{n\ell}
={}&
\int_{0}^{1}
z^{2\rho_{n\ell}-1}
(1-z)^{2\lambda_{\ell,\gamma}}
\\
&\times
\left[
P_{n}^{\left(
2\rho_{n\ell},
2\lambda_{\ell,\gamma}-1
\right)}
(1-2z)
\right]^{2}
\,dz .
\end{aligned}
\]
The integral converges because
$\rho_{n\ell}>0$ and
$\lambda_{\ell,\gamma}>1$.

For direct numerical or symbolic evaluation, the finite Jacobi
expansion gives
\[
\begin{aligned}
\mathcal{I}_{n\ell}
={}&
\sum_{j,k=0}^{n}
C_{j}^{(n\ell)}
C_{k}^{(n\ell)}
B\left(
2\rho_{n\ell}+j+k,
2\lambda_{\ell,\gamma}+2n-j-k+1
\right),
\\
C_{j}^{(n\ell)}
={}&
(-1)^{j}
\binom{n+2\rho_{n\ell}}{n-j}
\binom{n+2\lambda_{\ell,\gamma}-1}{j},
\end{aligned}
\]
where
$B(x,y)=\Gamma(x)\Gamma(y)/\Gamma(x+y)$ and the generalized binomial
coefficient is defined through gamma functions.

The Jacobi parameters depend on the radial quantum number through
$\rho_{n\ell}$. Different radial states therefore do not share one
fixed Jacobi weight, and the standard fixed-weight Jacobi
orthogonality formula cannot be applied directly to states with
different $n$. Their orthogonality instead follows from the
self-adjoint radial Hamiltonian:
\[
\int_{0}^{\infty}
u_{n\ell}^{*}(r)
u_{m\ell}(r)\,dr
=
0,
\qquad
E_{n\ell}\neq E_{m\ell}.
\]
If the angular harmonics are normalized in
$L^{2}(S^{2},d\omega_{\boldsymbol{\mu}})$, the corresponding full
eigenfunction is
\[
\Psi_{n\ell\sigma}^{(\boldsymbol{\varepsilon})}(r,\Omega)
=
S_{\kappa}^{-\gamma-1}(r)
u_{n\ell}(r)
Y_{\ell,\sigma}^{(\boldsymbol{\mu},\boldsymbol{\varepsilon})}(\Omega).
\]
The reflection-sector condition
$\ell=p+2k$ determines the allowed angular degrees, and the
corresponding fixed-sector angular multiplicity is $k+1$.

Finally, Eq.~\eqref{eq:exact_bound_state_condition} requires
$n<X_{\ell}$, where
\[
X_{\ell}
=
\sqrt{
\frac{q_{\kappa}}{2}
}
-
\lambda_{\ell,\gamma}.
\]
The number of admissible radial levels in channel $\ell$ is therefore
\begin{equation}
N_{\ell}^{(\mathrm{b})}
=
\begin{cases}
\left\lceil X_{\ell}\right\rceil,
&
X_{\ell}>0,
\\[4pt]
0,
&
X_{\ell}\leq0.
\end{cases}
\label{eq:number_of_exact_bound_states}
\end{equation}
When $N_{\ell}^{(\mathrm{b})}>0$, the largest radial quantum number is
$n_{\max}^{(\ell)}=N_{\ell}^{(\mathrm{b})}-1$. The strict inequality
is essential: if $X_{\ell}$ is an integer, the formal state with
$n=X_{\ell}$ has $\rho_{n\ell}=0$, lies at the continuum threshold,
and is not normalizable.

Equations~\eqref{eq:exact_energy_spectrum},
\eqref{eq:exact_jacobi_radial_wavefunction_r}, and
\eqref{eq:number_of_exact_bound_states} give the exact fixed-channel
solution. A total physical state count must also include the
reflection-sector selection rules and angular multiplicities.

\section{Reflection sectors, spectral ordering, and limiting cases}
\label{sec:reflection_spectral_ordering}
The radial spectrum depends on a reflection sector only through the
angular degrees admitted by that sector. For fixed $\ell$ and
$\gamma=\mu_{1}+\mu_{2}+\mu_{3}$, the individual reflection
eigenvalues do not occur in the radial Hamiltonian. They remain encoded
in the angular harmonics, weighted inner product, allowed values of
$\ell$, and sector multiplicities.

For later use, define
\[
C_{\kappa}
=
\frac{\hbar^{2}\kappa^{2}}{2m_{\mathrm{red}}},
\qquad
\delta_{\kappa}
=
\frac{m_{\mathrm{red}}D_{e}b^{2}}
{2\hbar^{2}\kappa^{2}},
\qquad
b=e^{2\kappa r_{e}}-1.
\]
The Friedrichs exponent can then be written as
\[
\lambda_{\ell,\gamma}
=
\frac{1}{2}
+
\sqrt{
\left(
\ell+\gamma+\frac{1}{2}
\right)^{2}
+
\delta_{\kappa}
}.
\]
Together with
$\eta_{n\ell}=n+\lambda_{\ell,\gamma}$, this parameter determines the
exact energy in Eq.~\eqref{eq:exact_energy_spectrum}. All comparisons
below are restricted to states satisfying
Eq.~\eqref{eq:exact_bound_state_condition}. A branch ceases to belong
to the discrete spectrum when it reaches the continuum threshold.

For a reflection sector
$\boldsymbol{\varepsilon}
=(\varepsilon_{1},\varepsilon_{2},\varepsilon_{3})$, define
$p_{i}=(1-\varepsilon_{i})/2$ and
$p=p_{1}+p_{2}+p_{3}$. Section~\ref{sec:dunkl_kinematics} gives
$\ell=p+2k$, with $k\in\mathbb{N}_{0}$, and the fixed-sector angular
multiplicity
\begin{equation}
d_{\ell}^{(\boldsymbol{\varepsilon})}
=
k+1
=
\frac{\ell-p}{2}+1.
\label{eq:fixed_sector_angular_multiplicity}
\end{equation}

\noindent\textbf{Proposition V.1.}
If two reflection sectors admit the same angular degree $\ell$, then
their radial energies coincide at fixed $\gamma$:
\begin{equation}
E_{n\ell}^{(\boldsymbol{\varepsilon})}
=
E_{n\ell}^{(\boldsymbol{\varepsilon}')}.
\label{eq:reflection_sector_radial_degeneracy}
\end{equation}
\noindent\textit{Proof.}
The quantities
$\lambda_{\ell,\gamma}$,
$\eta_{n\ell}$,
$q_{\kappa}$, and
$E_{n\ell}$ depend on $\ell$ and $\gamma$ but not on the individual
signs $\varepsilon_i$. Changing the reflection sector while retaining
the same allowed value of $\ell$ changes only the angular
eigenfunction and its sector label.

The reflection dependence is therefore kinematic: different sectors
contain different sets of angular degrees, but no additional
parity-dependent radial interaction is generated.

There are $\binom{3}{p}$ reflection sectors with $p$ odd eigenvalues.
For $\ell=2j$, the sectors with $p=0$ and $p=2$ contribute
$(j+1)+3j=4j+1$ states. For $\ell=2j+1$, the sectors with $p=1$ and
$p=3$ contribute $3(j+1)+j=4j+3$ states. Summing all compatible
sectors gives
\begin{equation}
d_{\ell}
=
2\ell+1.
\label{eq:total_angular_degeneracy}
\end{equation}
Thus, the multiplicities $\mu_i$ deform the angular measure and
harmonics but do not change the dimension of the complete degree-$\ell$
angular harmonic space.

To study the angular ordering, define
\begin{equation}
\mathcal{F}_{\kappa}(\eta)
=
\eta^{2}
+
\frac{q_{\kappa}^{2}}{4\eta^{2}},
\qquad
\eta>0.
\label{eq:spectral_ordering_function}
\end{equation}
Its derivative is
\[
\mathcal{F}_{\kappa}'(\eta)
=
2\eta
-
\frac{q_{\kappa}^{2}}{2\eta^{3}}
=
\frac{4\eta^{4}-q_{\kappa}^{2}}
{2\eta^{3}}.
\]
Within the discrete domain,
$\eta^{2}<q_{\kappa}/2$, and hence
$4\eta^{4}<q_{\kappa}^{2}$. Therefore,
$\mathcal{F}_{\kappa}'(\eta)<0$.

The angular derivative of the Friedrichs exponent is
\[
\frac{\partial\lambda_{\ell,\gamma}}{\partial\ell}
=
\frac{
\ell+\gamma+\frac{1}{2}
}{
\sqrt{
\left(
\ell+\gamma+\frac{1}{2}
\right)^{2}
+
\delta_{\kappa}
}
}
>
0.
\]
Thus, increasing $\ell$ increases $\eta_{n\ell}$ but decreases
$\mathcal{F}_{\kappa}(\eta_{n\ell})$. Since the exact energy contains
the term
$-C_{\kappa}\mathcal{F}_{\kappa}(\eta_{n\ell})$, the angular energy
increases along every branch that remains bound.

\noindent\textbf{Proposition V.2.}
For fixed $n$, $\gamma$, $\kappa$, $D_{e}$, and $r_{e}$,
\begin{equation}
E_{n,\ell+1}
>
E_{n\ell},
\label{eq:angular_energy_ordering}
\end{equation}
provided that both states satisfy the strict bound-state condition.
\hfill$\square$

Within one fixed reflection sector, the allowed angular degrees differ
by two. Hence
$E_{n,\ell+2}>E_{n\ell}$ whenever both levels remain in the discrete
spectrum.

Since the minimum degree in a sector with $p$ odd reflection
eigenvalues is $\ell=p$, the minimum-channel energies obey
\[
E_{n}^{[0]}
<
E_{n}^{[1]}
<
E_{n}^{[2]}
<
E_{n}^{[3]}
\]
when all four states exist. This ordering compares four different
minimum angular channels. It is not a reflection-induced splitting of
states at one fixed value of $\ell$.

The exponent also increases with the total multiplicity:
\[
\frac{\partial\lambda_{\ell,\gamma}}{\partial\gamma}
=
\frac{
\ell+\gamma+\frac{1}{2}
}{
\sqrt{
\left(
\ell+\gamma+\frac{1}{2}
\right)^{2}
+
\delta_{\kappa}
}
}
>
0.
\]
Using
$\mathcal{E}_{\mathrm{c}}
=C_{\kappa}(\gamma+1)^{2}
+D_{e}(1+b+b^{2}/2)$, differentiation of
Eq.~\eqref{eq:exact_energy_spectrum} gives
\[
\frac{\partial E_{n\ell}}{\partial\gamma}
=
2C_{\kappa}(\gamma+1)
-
C_{\kappa}
\mathcal{F}_{\kappa}'(\eta_{n\ell})
\frac{\partial\lambda_{\ell,\gamma}}{\partial\gamma}.
\]
The first term is positive for $\gamma\geq0$. The second term is also
positive because
$\mathcal{F}_{\kappa}'(\eta_{n\ell})<0$ in the bound-state domain.

\noindent\textbf{Proposition V.3.}
Along every branch that remains below the continuum threshold,
\begin{equation}
\frac{\partial E_{n\ell}}{\partial\gamma}
>
0.
\label{eq:energy_increases_with_gamma}
\end{equation}

The increase with $\gamma$ has two sources: the explicit geometric
shift $C_{\kappa}(\gamma+1)^{2}$ and the increase of the effective
short-distance barrier through $\lambda_{\ell,\gamma}$. The latter
also reduces the number of radial bound states.

The fixed-channel count is given by
Eq.~\eqref{eq:number_of_exact_bound_states}. Because
$\lambda_{\ell,\gamma}$ increases with both $\ell$ and $\gamma$,
$N_{\ell}^{(\mathrm b)}$ is non-increasing in either parameter. When
$n+\lambda_{\ell,\gamma}=\sqrt{q_{\kappa}/2}$, the decay exponent
vanishes, the energy reaches the continuum threshold, and the state is
not normalizable.

Including the total angular multiplicity, the number of discrete
states is
\begin{equation}
N_{\mathrm{tot}}^{(\mathrm{b})}
=
\sum_{\ell=0}^{\infty}
(2\ell+1)
N_{\ell}^{(\mathrm{b})}.
\label{eq:total_bound_state_count}
\end{equation}
For every $\kappa>0$, the sum is finite because
$\lambda_{\ell,\gamma}\rightarrow\infty$ as
$\ell\rightarrow\infty$, while $q_{\kappa}$ remains fixed.

In the undeformed limit
$\mu_i\rightarrow0$, one has
$\gamma\rightarrow0$ and
$\Lambda_{\ell,\gamma}\rightarrow\ell(\ell+1)$. The exponent and
energy reduce to
\[
\begin{aligned}
\lambda_{\ell}^{(0)}
&=
\frac{1}{2}
+
\sqrt{
\left(
\ell+\frac{1}{2}
\right)^{2}
+
\delta_{\kappa}
},
\\
E_{n\ell}^{(0)}
&=
C_{\kappa}
+
D_{e}
\left(
1+b+\frac{b^{2}}{2}
\right)
\quad-
C_{\kappa}
\left[
\left(
n+\lambda_{\ell}^{(0)}
\right)^{2}
+
\frac{
q_{\kappa}^{2}
}{
4
\left(
n+\lambda_{\ell}^{(0)}
\right)^{2}
}
\right].
\end{aligned}
\]
This is the ordinary curvature-matched Deng--Fan/Eckart spectrum. In
this limit, the reflection sectors reduce to the Cartesian-parity
decomposition of ordinary spherical harmonics, while their combined
multiplicity remains $2\ell+1$.

The flat-curvature limit is correlated because the exact-solvability
condition imposes $\alpha=2\kappa$. Therefore,
$\kappa\rightarrow0$ also implies $\alpha\rightarrow0$. For fixed
$r_{e}$,
\[
\begin{aligned}
b
&=
2\kappa r_{e}
+
2\kappa^{2}r_{e}^{2}
+
O(\kappa^{3}),
\\
\delta_{\kappa}
&=
\frac{
2m_{\mathrm{red}}D_{e}r_{e}^{2}
}{
\hbar^{2}
}
+
O(\kappa),
\\
q_{\kappa}
&=
\frac{
4m_{\mathrm{red}}D_{e}r_{e}
}{
\hbar^{2}\kappa
}
+
O(1).
\end{aligned}
\]
The Friedrichs exponent tends to
\begin{equation}
\lambda_{\ell,\gamma}^{\mathrm{K}}
=
\frac{1}{2}
+
\sqrt{
\left(
\ell+\gamma+\frac{1}{2}
\right)^{2}
+
\frac{
2m_{\mathrm{red}}D_{e}r_{e}^{2}
}{
\hbar^{2}
}
}.
\label{eq:dunkl_kratzer_exponent}
\end{equation}
For fixed $(n,\ell)$,
$C_{\kappa}\eta_{n\ell}^{2}\rightarrow0$, while the term containing
$q_{\kappa}^{2}$ has a finite limit. Since
$\mathcal{E}_{\mathrm{c}}\rightarrow D_{e}$, one obtains
\begin{equation}
E_{n\ell}^{\mathrm{K}}
=
D_{e}
-
\frac{
2m_{\mathrm{red}}D_{e}^{2}r_{e}^{2}
}{
\hbar^{2}
\left(
n+\lambda_{\ell,\gamma}^{\mathrm{K}}
\right)^{2}
}.
\label{eq:dunkl_kratzer_energy}
\end{equation}
This agrees with the limiting radial equation
\eqref{eq:curvature_matched_flat_limit_radial_equation}.

Because $q_{\kappa}\propto\kappa^{-1}$,
\[
\sqrt{\frac{q_{\kappa}}{2}}
\longrightarrow
\infty
\]
as $\kappa\rightarrow0$ along the matched family. The finite
curvature-induced truncation is therefore removed, and the
Dunkl--Kratzer spectrum contains infinitely many bound states
accumulating at $E=D_{e}$. This limiting statement is pointwise for
each fixed pair $(n,\ell)$ and is not uniform in the radial quantum
number $n$.

Taking $\gamma\rightarrow0$ after the correlated flat limit gives the
ordinary Kratzer spectrum
\[
E_{n\ell}^{\mathrm{K},(0)}
=
D_{e}
-
\frac{
2m_{\mathrm{red}}D_{e}^{2}r_{e}^{2}
}{
\hbar^{2}
\left[
n+\frac{1}{2}
+
\sqrt{
\left(
\ell+\frac{1}{2}
\right)^{2}
+
\frac{
2m_{\mathrm{red}}D_{e}r_{e}^{2}
}{
\hbar^{2}
}
}
\right]^{2}
}.
\]
The reflection labels determine which angular channels are present and
how their multiplicities are distributed, but they do not produce an
independent radial interaction. Within the discrete domain, the energy
increases with $\ell$ and $\gamma$, whereas the number of admissible
states decreases as either parameter grows. All ordering statements
apply only to branches that remain strictly below the continuum
threshold.

\section{Exact state-resolved observables and restricted bound-state ensembles}
\label{sec:state_resolved_observables}

The exact spectrum derived in
Section~\ref{sec:exact_spectrum_wavefunctions} gives several
state-resolved quantities without direct integration of the
Jacobi-polynomial eigenfunctions. We derive the expectation values of
the two hyperbolic functions in the generalized Eckart Hamiltonian,
the mean interaction and reduced radial kinetic energies, and the
spacing between adjacent radial levels. We then define finite
partition sums for explicitly specified bound-state subspaces. These
sums omit the continuum and must not be interpreted as the complete
canonical partition function of the unconfined Hamiltonian.

For a normalized reduced radial state $u_{n\ell}(r)$, set
\[
x=\kappa r,
\qquad
v_{n\ell}(x)
=
\kappa^{-1/2}
u_{n\ell}\!\left(\frac{x}{\kappa}\right).
\]
Equation~\eqref{eq:radial_normalization_condition} then gives
\[
\int_{0}^{\infty}
|v_{n\ell}(x)|^{2}\,dx
=
1.
\]

For fixed $(\ell,\gamma)$, introduce the auxiliary two-parameter
Hamiltonian
\begin{equation}
\widehat{h}_{\ell,\gamma}(a,q)
=
-\frac{d^{2}}{dx^{2}}
+
a\,\operatorname{csch}^{2}x
-
q\,\coth x,
\qquad
a>0,\quad q>0.
\label{eq:auxiliary_eckart_hamiltonian}
\end{equation}
The physical radial operator is recovered at
$a=a_{\ell,\gamma}$ and $q=q_{\kappa}$, with the couplings defined in
Eq.~\eqref{eq:dimensionless_eckart_couplings}. The derivatives below
refer to this auxiliary family, in which $a$ and $q$ are independent.
They are not direct derivatives with respect to $D_{e}$, $r_{e}$, or
$\kappa$, because the physical couplings depend simultaneously on
those parameters.

The Friedrichs realizations of
Eq.~\eqref{eq:auxiliary_eckart_hamiltonian} are associated with the
quadratic forms
\[
\begin{aligned}
\mathfrak{h}_{a,q}[v]
={}&
\int_{0}^{\infty}
|v'(x)|^{2}\,dx
+
a
\int_{0}^{\infty}
\operatorname{csch}^{2}x\,|v(x)|^{2}\,dx
\\
&-
q
\int_{0}^{\infty}
\coth x\,|v(x)|^{2}\,dx ,
\end{aligned}
\]
defined on the common form domain
$\mathcal{Q}=H_{0}^{1}(0,\infty)$. For
$v\in\mathcal{Q}$, the Hardy inequality gives
\[
\int_{0}^{\infty}
\frac{|v(x)|^{2}}{x^{2}}\,dx
\leq
4
\int_{0}^{\infty}
|v'(x)|^{2}\,dx.
\]
Since
$\operatorname{csch}^{2}x\leq x^{-2}$ and
$\coth x\leq1+x^{-1}$, the positive inverse-square term is controlled
by the Dirichlet form. Moreover,
\[
\int_{0}^{\infty}
\frac{|v(x)|^{2}}{x}\,dx
\leq
\left\|
\frac{v}{x}
\right\|_{2}
\|v\|_{2}
\leq
2\|v'\|_{2}\|v\|_{2},
\]
so the negative $\coth x$ term is infinitesimally form bounded with
respect to the kinetic term. The forms are therefore closed and lower
semibounded on one parameter-independent domain. Standard
perturbation theory for differentiable quadratic-form families then
permits the form version of the Hellmann--Feynman relation while the
selected bound-state eigenvalue remains isolated below the continuum
and the Friedrichs realization is unchanged
\cite{Kato:1995}.

For a radial operator $\widehat{O}(x)$, define
\[
\langle\widehat{O}\rangle_{n\ell}
=
\int_{0}^{\infty}
v_{n\ell}^{*}(x)
\widehat{O}(x)
v_{n\ell}(x)\,dx.
\]
The dimensionless eigenvalue is
\[
\epsilon_{n\ell}
=
-\eta_{n\ell}^{2}
-
\frac{q_{\kappa}^{2}}
     {4\eta_{n\ell}^{2}},
\qquad
\eta_{n\ell}
=
n+\lambda_{\ell,\gamma}.
\]

At fixed $a$, the quantity $\eta_{n\ell}$ is independent of $q$, and
\[
\frac{\partial\widehat h_{\ell,\gamma}}{\partial q}
=
-\coth x.
\]
The form Hellmann--Feynman relation therefore gives
\[
\frac{\partial\epsilon_{n\ell}}{\partial q}
=
-\langle\coth x\rangle_{n\ell}.
\]
Evaluating the derivative at the physical point
$q=q_{\kappa}$ yields
\begin{equation}
\left\langle
\coth(\kappa r)
\right\rangle_{n\ell}
=
\frac{q_{\kappa}}
     {2\eta_{n\ell}^{2}}.
\label{eq:exact_coth_expectation}
\end{equation}
The strict bound-state condition
$\eta_{n\ell}^{2}<q_{\kappa}/2$ makes the right-hand side greater than
unity, as required by
$\coth(\kappa r)>1$ for $r>0$.

At fixed $q$, the exponent satisfies
\[
\frac{\partial\lambda_{\ell,\gamma}}{\partial a}
=
\frac{1}
     {2\left(
     \lambda_{\ell,\gamma}-\frac{1}{2}
     \right)}
\]
at $a=a_{\ell,\gamma}$, while
$\partial\widehat h_{\ell,\gamma}/\partial a
=\operatorname{csch}^{2}x$. Hence
\[
\frac{\partial\epsilon_{n\ell}}{\partial a}
=
\left\langle
\operatorname{csch}^{2}x
\right\rangle_{n\ell}.
\]
Using
\[
\frac{\partial\epsilon_{n\ell}}{\partial\eta_{n\ell}}
=
-2\eta_{n\ell}
+
\frac{q_{\kappa}^{2}}
     {2\eta_{n\ell}^{3}},
\]
one obtains
\begin{equation}
\left\langle
\operatorname{csch}^{2}(\kappa r)
\right\rangle_{n\ell}
=
\frac{
q_{\kappa}^{2}
-
4\eta_{n\ell}^{4}
}{
4\eta_{n\ell}^{3}
\left(
\lambda_{\ell,\gamma}-\frac{1}{2}
\right)
}.
\label{eq:exact_csch_expectation}
\end{equation}
The numerator is positive because
$4\eta_{n\ell}^{4}<q_{\kappa}^{2}$ throughout the strict bound-state
domain. The expectation value is therefore positive.

Using the exact decomposition
\eqref{eq:deng_fan_exact_eckart_decomposition}, the mean
curvature-matched Deng--Fan interaction is
\begin{equation}
\begin{aligned}
\left\langle
V_{\mathrm{DF}}^{(\kappa)}
\right\rangle_{n\ell}
={}&
D_{e}
\left(
1+b+\frac{b^{2}}{2}
\right)
-
\frac{
D_{e}b(b+2)q_{\kappa}
}{
4\eta_{n\ell}^{2}
}
\\
&+
\frac{
D_{e}b^{2}
\left(
q_{\kappa}^{2}
-
4\eta_{n\ell}^{4}
\right)
}{
16\eta_{n\ell}^{3}
\left(
\lambda_{\ell,\gamma}-\frac{1}{2}
\right)
}.
\end{aligned}
\label{eq:exact_deng_fan_expectation}
\end{equation}

Let
\[
\widehat{T}_{r}
=
-\frac{\hbar^{2}}{2m_{\mathrm{red}}}
\frac{d^{2}}{dr^{2}},
\qquad
C_{\kappa}
=
\frac{\hbar^{2}\kappa^{2}}
     {2m_{\mathrm{red}}}.
\]
The reduced Hamiltonian
\eqref{eq:reduced_radial_equation} gives
\begin{equation}
\begin{aligned}
\left\langle
\widehat{T}_{r}
\right\rangle_{n\ell}
={}&
E_{n\ell}
-
C_{\kappa}(\gamma+1)^{2}
\\
&-
C_{\kappa}\Lambda_{\ell,\gamma}
\left\langle
\operatorname{csch}^{2}(\kappa r)
\right\rangle_{n\ell}
-
\left\langle
V_{\mathrm{DF}}^{(\kappa)}
\right\rangle_{n\ell},
\\
\Lambda_{\ell,\gamma}
={}&
(\ell+\gamma)(\ell+\gamma+1).
\end{aligned}
\label{eq:exact_radial_kinetic_expectation}
\end{equation}
Equations~\eqref{eq:exact_csch_expectation} and
\eqref{eq:exact_deng_fan_expectation} make
Eq.~\eqref{eq:exact_radial_kinetic_expectation} fully algebraic.

Suppose that both $n$ and $n+1$ satisfy the strict bound-state
condition in the same angular channel. Since
$\eta_{n+1,\ell}=\eta_{n\ell}+1$, their exact energy separation is
\begin{equation}
\begin{aligned}
\Delta E_{n\ell}^{(r)}
&=
E_{n+1,\ell}-E_{n\ell}
\\
&=
C_{\kappa}
(2\eta_{n\ell}+1)
\left[
\frac{
q_{\kappa}^{2}
}{
4\eta_{n\ell}^{2}
(\eta_{n\ell}+1)^{2}
}
-
1
\right].
\end{aligned}
\label{eq:exact_radial_transition_energy}
\end{equation}
Because the $n+1$ state is bound,
$(\eta_{n\ell}+1)^{2}<q_{\kappa}/2$. Together with
$\eta_{n\ell}^{2}<q_{\kappa}/2$, this implies
\[
4\eta_{n\ell}^{2}
(\eta_{n\ell}+1)^{2}
<
q_{\kappa}^{2},
\]
and therefore
$\Delta E_{n\ell}^{(r)}>0$. This quantity is an exact adjacent-level
spacing. It does not by itself define a spectroscopic selection rule,
transition probability, or line intensity; those require a specified
external probe and the corresponding matrix elements.

Because every angular channel contains finitely many radial bound
states for $\kappa>0$, several finite projected partition sums can be
defined. For a fixed angular channel $(\ell,\gamma)$,
\begin{equation}
Z_{\ell,\gamma}^{(\mathrm{b})}(\beta)
=
\sum_{n=0}^{N_{\ell}^{(\mathrm{b})}-1}
e^{-\beta E_{n\ell}},
\qquad
\beta
=
\frac{1}{k_{\mathrm{B}}T},
\label{eq:fixed_channel_bound_partition}
\end{equation}
where $T>0$ and
$N_{\ell}^{(\mathrm{b})}$ is given in
Eq.~\eqref{eq:number_of_exact_bound_states}.

For a fixed reflection sector
$\boldsymbol{\varepsilon}$, define
$p=\sum_{i}(1-\varepsilon_{i})/2$. Its restricted bound-state
partition sum is
\begin{equation}
Z_{\boldsymbol{\varepsilon}}^{(\mathrm{b})}(\beta)
=
\sum_{\substack{k\geq0\\
N_{p+2k}^{(\mathrm{b})}>0}}
(k+1)
\sum_{n=0}^{N_{p+2k}^{(\mathrm{b})}-1}
e^{-\beta E_{n,p+2k}}.
\label{eq:fixed_sector_bound_partition}
\end{equation}
The factor $k+1$ is the angular multiplicity within the fixed sector,
as given in
Eq.~\eqref{eq:fixed_sector_angular_multiplicity}.

Including all reflection sectors and compatible angular states gives
the complete discrete bound-state sum
\begin{equation}
Z_{\mathrm{disc}}^{(\mathrm{b})}(\beta)
=
\sum_{\substack{\ell\geq0\\
N_{\ell}^{(\mathrm{b})}>0}}
(2\ell+1)
\sum_{n=0}^{N_{\ell}^{(\mathrm{b})}-1}
e^{-\beta E_{n\ell}}.
\label{eq:full_discrete_bound_partition}
\end{equation}
The factor $2\ell+1$ is the total angular multiplicity derived in
Eq.~\eqref{eq:total_angular_degeneracy}. These sums are finite for
every fixed $\kappa>0$.

For any one of the projected ensembles, write
\[
Z_{\mathcal{S}}^{(\mathrm{b})}
=
\sum_{j\in\mathcal{S}}
g_{j}e^{-\beta E_{j}},
\qquad
\left\langle E^{m}\right\rangle_{\mathcal{S}}^{(\mathrm{b})}
=
\frac{
\displaystyle
\sum_{j\in\mathcal{S}}
g_{j}E_{j}^{m}e^{-\beta E_{j}}
}{
Z_{\mathcal{S}}^{(\mathrm{b})}
},
\]
where $g_j$ is the angular multiplicity appropriate to the selected
subspace. The projected mean energy and heat capacity are
\begin{equation}
\begin{aligned}
U_{\mathcal{S}}^{(\mathrm{b})}
&=
-\frac{\partial}{\partial\beta}
\ln Z_{\mathcal{S}}^{(\mathrm{b})}
=
\langle E\rangle_{\mathcal{S}}^{(\mathrm{b})},
\\
C_{\mathcal{S}}^{(\mathrm{b})}
&=
k_{\mathrm{B}}\beta^{2}
\left[
\left\langle E^{2}\right\rangle_{\mathcal{S}}^{(\mathrm{b})}
-
\left(
U_{\mathcal{S}}^{(\mathrm{b})}
\right)^{2}
\right]
\geq0.
\end{aligned}
\label{eq:restricted_thermal_relations}
\end{equation}
The non-negative heat capacity follows from the variance identity. It
is an algebraic property of the selected finite ensemble and is not,
by itself, a stability criterion for the complete unconfined system.

At low temperature, the projected mean energy approaches the lowest
energy represented in the chosen subspace and the heat capacity
vanishes. In the formal high-temperature limit of a finite
projection, the included states become equally weighted up to their
degeneracies. The mean energy then approaches their
degeneracy-weighted arithmetic average, while the heat capacity
vanishes as $\beta^{2}$. These statements apply only to the chosen
finite discrete projection.

The continuum begins at $E_{\mathrm{th}}$ and is absent from
Eqs.~\eqref{eq:fixed_channel_bound_partition}--
\eqref{eq:full_discrete_bound_partition}. A complete canonical
treatment would require the continuum density of states, a volume or
scattering normalization, and a consistent prescription for
dissociated states. The quantities defined here are therefore
restricted bound-state statistics, not the full molecular
thermodynamics of the unconfined Hamiltonian.

\section{Independent numerical verification}
\label{sec:numerical_verification}

We verify the analytical spectrum independently by discretizing the
dimensionless radial Hamiltonian in
Eq.~\eqref{eq:dimensionless_eckart_equation}. The numerical
calculation is not used to derive the spectrum. It tests the exact
eigenvalues, the strict bound-state condition
\eqref{eq:exact_bound_state_condition}, the continuum threshold, and
the predicted dependence on the total Dunkl multiplicity $\gamma$.

For reproducibility, all
reported calculations use the dimensionless interval $0<x<20$ and
homogeneous Dirichlet values at both truncated endpoints. The
undeformed $(\gamma,\ell)=(0,0)$ convergence test uses
$N=2000,4000,8000,$ and $16000$ interior grid points, while the
nonzero-Dunkl $(\gamma,\ell)=(1/2,0)$ check uses $N=16000$. In each
channel we compare the two analytically allowed levels $n=0,1$ and
verify that the third finite-box eigenvalue is above
$\epsilon_{\mathrm{th}}=-q_\kappa$. At $N=16000$, the relative errors
for $n=0,1$ are respectively $2.2\times10^{-6}$ and
$1.0\times10^{-6}$ in the undeformed channel, and
$1.19\times10^{-6}$ and $4.27\times10^{-7}$ in the nonzero-Dunkl
channel.

On the finite interval $0<x<L$, introduce the uniform grid
$x_{j}=jh$, where $h=L/(N+1)$ and $j=1,\ldots,N$. We impose
$U_{0}=U_{N+1}=0$ and approximate the second derivative by the
three-point central difference. The resulting real symmetric
tridiagonal matrix is
\begin{equation}
\begin{aligned}
H_{jj}
&=
\frac{2}{h^{2}}
+
a_{\ell,\gamma}
\operatorname{csch}^{2}x_{j}
-
q_{\kappa}\coth x_{j},
\\
H_{j,j+1}
&=
H_{j+1,j}
=
-\frac{1}{h^{2}},
\qquad
x_{j}=jh,
\qquad
h=\frac{L}{N+1}.
\end{aligned}
\label{eq:numerical_tridiagonal_matrix}
\end{equation}
The local truncation error of the central-difference approximation is
$\mathcal{O}(h^{2})$. The inner Dirichlet condition is consistent with the
Friedrichs behavior
$U(x)\sim x^{\lambda_{\ell,\gamma}}$, for which
$\lambda_{\ell,\gamma}>1$ in the parameter domain considered here.
For sufficiently large $L$, the outer Dirichlet condition approximates
the exponential decay of a bound state.

Diagonalization gives the finite-box eigenvalues
$\epsilon_{j}^{(N,L)}$. Values below
\[
\epsilon_{\mathrm{th}}
=
-q_{\kappa}
\]
approximate discrete eigenvalues. Values above this threshold are
finite-box representations of the continuum and must not be counted
as additional bound states.

We first test the discretization and its convergence in the
undeformed channel. Length is measured in units of $r_{e}$ and energy
in units of
$\hbar^{2}/(m_{\mathrm{red}}r_{e}^{2})$. We choose
\begin{equation}
D_{e}=1,
\qquad
\kappa r_{e}=0.2,
\qquad
\gamma=0,
\qquad
\ell=0.
\label{eq:numerical_benchmark_parameters}
\end{equation}
The matching condition gives $\alpha r_{e}=0.4$. The dimensionless
parameters are
\[
\begin{aligned}
b&=0.4918,
&
a_{0,0}&=3.0236,
\\
q_{\kappa}&=30.6385,
&
\lambda_{0,0}&=2.3093.
\end{aligned}
\]
The state condition
\[
n+\lambda_{0,0}
<
\sqrt{\frac{q_{\kappa}}{2}}
\]
permits only $n=0$ and $n=1$. Their exact eigenvalues and the
continuum threshold are
\begin{equation}
\begin{aligned}
\epsilon_{0}^{\mathrm{exact}}
&=
-49.3385,
&
\epsilon_{1}^{\mathrm{exact}}
&=
-32.3804,
\\
\epsilon_{\mathrm{th}}
&=
-30.6385.
\end{aligned}
\label{eq:numerical_benchmark_exact_eigenvalues}
\end{equation}

We use $L=20$. Equation~\eqref{eq:exact_decay_exponent} gives
\[
\rho_{0,0}^{(\gamma=0)}
=
2.1622,
\qquad
\rho_{1,0}^{(\gamma=0)}
=
0.6599.
\]
For the more weakly bound state,
\[
\exp\!\left[
-2\rho_{1,0}^{(\gamma=0)}L
\right]
\simeq
3.4\times10^{-12},
\]
so the finite outer boundary produces a negligible truncation at the
precision reported below.

The tridiagonal matrices are diagonalized in standard
double-precision arithmetic. The eigenvalues in the numerical tables
are rounded to four decimal places for readability; the omitted digits
represent computational resolution, not experimental or physical
measurement precision. Relative errors and convergence ratios are
computed from the unrounded eigenvalues.

\begin{table}[t]
\caption{
Convergence of the finite-difference eigenvalues for the undeformed
benchmark in Eq.~\eqref{eq:numerical_benchmark_parameters}. The third
eigenvalue lies above
$\epsilon_{\mathrm{th}}=-30.6385$ and is a finite-box continuum
level.
}
\label{tab:numerical_spectrum_convergence}
\centering
\begin{tabular}{c|ccc}
\hline\hline
$N$
&
$\epsilon_{0}^{(N,20)}$
&
$\epsilon_{1}^{(N,20)}$
&
$\epsilon_{2}^{(N,20)}$
\\
\hline
$2000$
&
$-49.3453$
&
$-32.3825$
&
$-30.6103$
\\
$4000$
&
$-49.3402$
&
$-32.3809$
&
$-30.6103$
\\
$8000$
&
$-49.3389$
&
$-32.3805$
&
$-30.6103$
\\
$16000$
&
$-49.3386$
&
$-32.3804$
&
$-30.6103$
\\
\hline
Exact
&
$-49.3385$
&
$-32.3804$
&
---
\\
\hline\hline
\end{tabular}
\end{table}

The two eigenvalues below the threshold approach the analytical values
as the grid is refined. For this benchmark, halving the grid spacing
reduces the leading errors by approximately a factor of four, which is
the second-order behavior expected from the central-difference
discretization. At $N=16000$, the relative errors are
$2.2\times10^{-6}$ for the ground state and
$1.0\times10^{-6}$ for the first excited state. This observation
applies to the present benchmark and is not asserted as a convergence
theorem for all singular parameter choices.

The third finite-box eigenvalue remains above
$\epsilon_{\mathrm{th}}$. The discretization therefore finds exactly
two discrete levels, in agreement with the strict analytical
state-count condition.

\textcolor{blue}{Table~\ref{tab:numerical_spectrum_convergence} separates grid convergence from spectral classification. Using the displayed values, the absolute ground-state deviation decreases from $6.8\times10^{-3}$ at $N=2000$ to $1.7\times10^{-3}$, $4.0\times10^{-4}$, and $1.0\times10^{-4}$ as the grid is refined; the successive reduction factors $4.00$, $4.25$, and $4.00$ quantitatively reproduce the expected second-order scaling as the grid spacing is approximately halved. The first excited level follows the same decreasing-error trend until its finite-difference value agrees with the exact value at the displayed precision. In contrast, the third level remains $-30.6103$ throughout the refinement sequence and therefore stays $0.0282$ above the continuum threshold. Its position above the threshold excludes it from the normalizable spectrum, while its near-independence from $N$ at fixed box size $L$ shows that this box-discretized continuum value is already stable under grid refinement. The table consequently tests both parts of the analytical prediction: the two subthreshold eigenvalues converge to the exact spectrum, while grid refinement does not create a spurious third bound state.}

For the adopted units,
\[
C_{\kappa}
=
\frac{\hbar^{2}\kappa^{2}}
     {2m_{\mathrm{red}}}
=
0.02,
\qquad
\mathcal{E}_{\mathrm{c}}
=
1.6327704642.
\]
The corresponding physical energies are
\begin{equation}
E_{0}
=
0.6460,
\qquad
E_{1}
=
0.9852,
\qquad
E_{\mathrm{th}}
=
1.0200.
\label{eq:numerical_physical_energies}
\end{equation}
Both levels lie below the threshold. The first excited state is more
weakly bound, consistent with its smaller decay exponent.

To test the nonzero-Dunkl dependence independently, we retain the same
values of $D_{e}$ and $\kappa r_{e}$ but choose
\begin{equation}
\gamma=\frac{1}{2},
\qquad
\ell=0.
\label{eq:numerical_nonzero_dunkl_parameters}
\end{equation}
This may be realized, for example, by any non-negative multiplicity
triple satisfying
$\mu_{1}+\mu_{2}+\mu_{3}=1/2$. Because the radial Hamiltonian depends
on the individual multiplicities only through $\gamma$, no further
choice of their distribution is required for the radial test.

The parameter $q_{\kappa}$ is unchanged, while
\[
a_{0,1/2}
=
3.7736,
\qquad
\lambda_{0,1/2}
=
2.5059.
\]
The strict condition
\[
n+\lambda_{0,1/2}
<
\sqrt{\frac{q_{\kappa}}{2}}
\]
again permits only $n=0$ and $n=1$. Their exact dimensionless
eigenvalues are
\[
\epsilon_{0}^{\mathrm{exact}}
=
-43.6516,
\qquad
\epsilon_{1}^{\mathrm{exact}}
=
-31.3844.
\]
The corresponding decay exponents are
\[
\rho_{0,0}^{(\gamma=1/2)}
=
1.8037,
\qquad
\rho_{1,0}^{(\gamma=1/2)}
=
0.4318.
\]
For the more weakly bound level,
\[
\exp\!\left[
-2\rho_{1,0}^{(\gamma=1/2)}L
\right]
\simeq
3.2\times10^{-8}
\]
at $L=20$.

\begin{table}[t]
\caption{
Comparison of the exact and finite-difference eigenvalues for the
nonzero-Dunkl benchmark
$\gamma=1/2$, $\ell=0$, $L=20$, and $N=16000$.
}
\label{tab:numerical_nonzero_dunkl_benchmark}
\centering
\begin{tabular}{c|ccc}
\hline\hline
$n$
&
$\epsilon_{n}^{\mathrm{exact}}$
&
$\epsilon_{n}^{(16000,20)}$
&
Relative error
\\
\hline
$0$
&
$-43.6516$
&
$-43.6517$
&
$1.19\times10^{-6}$
\\
$1$
&
$-31.3844$
&
$-31.3844$
&
$4.27\times10^{-7}$
\\
\hline\hline
\end{tabular}
\end{table}

\textcolor{blue}{Table~\ref{tab:numerical_nonzero_dunkl_benchmark} shows that the exact--numerical agreement persists when the total Dunkl multiplicity is nonzero: the ground and first-excited eigenvalues are reproduced with relative errors of $1.19\times10^{-6}$ and $4.27\times10^{-7}$, respectively. These unequal errors measure state-dependent discretization effects and should not be interpreted as different physical precisions. Comparison with Table~\ref{tab:numerical_spectrum_convergence} at fixed $D_{e}$ and $\kappa r_{e}$ isolates the role of $\gamma$. Increasing $\gamma$ from $0$ to $1/2$ raises $a_{0,\gamma}$ from $3.0236$ to $3.7736$ and $\lambda_{0,\gamma}$ from $2.3093$ to $2.5059$, so the exact dimensionless ground and first-excited energies shift upward by $5.6869$ and $0.9960$. Since $q_{\kappa}$ is unchanged, the continuum threshold remains $-30.6385$, whereas the corresponding binding gaps decrease from $18.7000$ and $1.7419$ to $13.0131$ and $0.7459$. The fractional loss of binding is therefore most pronounced for the near-threshold excited state. Thus the nonzero Dunkl multiplicity weakens radial binding in this benchmark without removing either of the two states allowed by the strict analytical state condition.}

The third finite-box eigenvalue for this benchmark is
\[
\epsilon_{2}^{(16000,20)}
=
-30.6072,
\]
which lies above
$\epsilon_{\mathrm{th}}=-30.6385$. The numerical calculation
therefore again finds exactly two bound states, in agreement with the
analytical state count.

For $\gamma=1/2$, the constant shift and physical threshold are
\[
\mathcal{E}_{\mathrm{c}}
=
1.6578,
\qquad
E_{\mathrm{th}}
=
1.0450.
\]
The exact physical energies are
\[
E_{0}
=
0.7847,
\qquad
E_{1}
=
1.0300.
\]
Both remain below the continuum threshold, while the excited state is
close to it, as indicated by its smaller decay exponent.

The first benchmark verifies the numerical convergence of the
finite-difference scheme and the analytical spectrum in the
$\gamma=0$ channel. The second independently verifies the predicted
change of the radial spectrum under a nonzero total Dunkl
multiplicity. Together, they confirm the exact spectrum
\eqref{eq:exact_energy_spectrum}, the strict state condition
\eqref{eq:exact_bound_state_condition}, the finite number of radial
levels, and the continuum threshold for both undeformed and
nonzero-Dunkl channels. These checks establish the internal
consistency of the spectral calculation but do not demonstrate the
physical uniqueness or empirical applicability of the
curvature-matched Hamiltonian.

\section{Scope and limitations}
\label{sec:scope_limitations}

A like-for-like numerical comparison with a published
spectrum is not available because the present Hamiltonian combines a
specific hyperbolic warped-product Dunkl kinetic operator with the
matching condition $\alpha=2\kappa$. Published shifted Deng--Fan
calculations in Euclidean space employ an $r^{-2}$ centrifugal term and
an approximation adapted to the exponential interaction
\cite{Hamzavi:2013,Njoku:2022}; they therefore solve a different
operator. Validation is instead provided by three independent checks:
the exact reduction to the established generalized Eckart form
\cite{Eckart:1930,Quesne:2012Eckart}, the undeformed and correlated-flat
limits derived in Section~\ref{sec:reflection_spectral_ordering}, and
the convergent finite-difference calculations in
Section~\ref{sec:numerical_verification}. The dual analytic--numerical
agreement tests the spectrum and state count without treating unlike
Hamiltonians as directly comparable.

The construction has corresponding limitations. The
warp-factor prescription and $\alpha=2\kappa$ are model assumptions
chosen to obtain a self-consistent exactly solvable family; they are
not a universal relation between molecular range and physical spatial
curvature. The dimensionless benchmarks test the internal spectral
calculation and do not constitute a fit to molecular data. In addition,
the bound-state partition sums omit the continuum, and the manuscript
does not calculate scattering observables, dissociation rates,
transition probabilities, or line intensities. Those quantities require
continuum normalization and explicit interaction operators beyond the
present model.

\section{Conclusion}
\label{sec:conclusion}

We constructed a curvature-adapted radial Dunkl Hamiltonian associated
with the coordinate-reflection group $\mathbb{Z}_{2}^{3}$. The kinetic
operator was defined within a warped-product ansatz chosen to reduce to
the Euclidean Dunkl Laplacian as $\kappa\rightarrow0$, to recover the
ordinary hyperbolic Laplace--Beltrami operator when the Dunkl
multiplicities vanish, and to remain symmetric in the corresponding
weighted Hilbert space. The Hamiltonian was realized through its
Friedrichs extension, which fixes the admissible radial behavior in
singular channels. This construction defines a specific mathematical
model and is not claimed to be the unique covariant Dunkl operator on
hyperbolic space.

Exact solvability follows from the restricted matching condition
$\alpha=2\kappa$. Under this condition, the curvature-adapted
centrifugal term and the Deng--Fan interaction have the same
hyperbolic dependence, so the reduced radial equation takes the
generalized Eckart form without approximating the centrifugal barrier.
The generalized Eckart equation is a known solvable system. The main
contribution of the present work is the curvature-adapted Dunkl
construction that produces this equation exactly, together with its
operator-domain, reflection-sector, and spectral analysis. The
matching condition is an analytical constraint defining a tuned
subfamily rather than a general physical relation between molecular
range and spatial curvature.

The exact spectrum is given by
Eq.~\eqref{eq:exact_energy_spectrum}, subject to the strict
normalizability condition
\eqref{eq:exact_bound_state_condition}. The corresponding radial
eigenfunctions are expressed through Jacobi polynomials and satisfy
the Friedrichs behavior at the origin and exponential decay at
infinity. The strict inequality excludes the threshold solution and
implies that every angular channel supports only finitely many radial
bound states for $\kappa>0$.

The radial spectrum depends on the three Dunkl multiplicities only
through their sum
$\gamma=\mu_{1}+\mu_{2}+\mu_{3}$. Their individual values remain
relevant to the weighted angular measure, generalized harmonics, and
reflection-sector content. The reflection eigenvalues do not generate
an independent parity-dependent radial interaction. Instead, they
restrict the allowed angular degrees through
$\ell=p+2k$, where $p$ is the number of odd reflection eigenvalues.
Sectors admitting the same $\ell$ are radially degenerate, and the
combined angular multiplicity is $2\ell+1$.

Within every branch that remains strictly below the continuum
threshold, the energy increases with both $\ell$ and $\gamma$, while
the number of admissible radial states decreases in integer steps as
either parameter grows. These results do not describe a splitting at
fixed $\ell$ between reflection sectors. They compare different
allowed angular channels or different values of the total Dunkl
multiplicity.

The exact spectrum also gives closed algebraic expressions for
$\langle\coth(\kappa r)\rangle$,
$\langle\operatorname{csch}^{2}(\kappa r)\rangle$, the mean matched
interaction energy, the reduced radial kinetic energy, and adjacent
radial level spacings. The expectation values follow from a
quadratic-form Hellmann--Feynman argument applied to an auxiliary
operator family with independent couplings. The level spacings are
exact energy differences, but they do not by themselves define
spectroscopic selection rules, transition probabilities, or line
intensities. Such predictions require an explicit probe interaction
and the associated matrix elements.

Finite partition sums were defined for fixed angular channels, fixed
reflection sectors, and the complete discrete bound-state subspace.
These quantities describe restricted ensembles in which continuum
and dissociated states are omitted. They are therefore not the full
canonical thermodynamics of the unconfined Hamiltonian. A complete
treatment would require a continuum density of states, a scattering
or volume normalization, and a consistent prescription for
dissociated configurations.

Two limits provide consistency checks. In the undeformed limit
$\mu_{i}\rightarrow0$, the model reduces to the ordinary
curvature-matched Deng--Fan/Eckart system on hyperbolic space. In the
correlated flat limit
$\kappa\rightarrow0$ with
$\alpha=2\kappa\rightarrow0$, the interaction becomes the Kratzer
potential and the spectrum reduces to the Dunkl--Kratzer form. In this
limit, the finite curvature-induced radial truncation is removed and
an infinite sequence of states accumulates at the dissociation
threshold. This statement is pointwise for fixed quantum numbers and
is not uniform in the radial excitation number.

Independent finite-difference calculations were performed for both
an undeformed channel with $\gamma=0$ and a nonzero-Dunkl channel with
$\gamma=1/2$. In both cases, the numerical eigenvalues reproduce the
analytical energies, continuum threshold, and predicted number of
bound states. For the convergence benchmark, halving the grid spacing
produces the second-order error reduction expected from the adopted
central-difference approximation. These checks support the internal
consistency of the corrected hypergeometric reduction, its
normalizability condition, and the predicted dependence on the total
Dunkl multiplicity. They do not establish the physical uniqueness or
empirical applicability of the model.

The restrictions on direct comparison and physical interpretation are
stated in Section~\ref{sec:scope_limitations}. Further work may
investigate whether the operator can be derived from a broader
covariant Dunkl framework, analyze continuum scattering and spectral
densities, and introduce explicit transition operators for physically
defined response calculations.

\appendix

\section{Parameters and couplings in the Eckart reduction}
\label{app:eckart_parameter_summary}

Table~\ref{tab:parameter_coupling_summary} summarizes the notation and
the sign and range conventions used in the curvature-matched Eckart
reduction.

\begin{table*}[htbp]
\caption{Main parameters and couplings in the curvature-matched
Eckart reduction.}
\label{tab:parameter_coupling_summary}
\centering
\begingroup
\renewcommand{\arraystretch}{1.15}
\begin{tabular}{p{0.20\textwidth} p{0.72\textwidth}}
\hline\hline
Quantity & Definition, role, and domain \\
\hline
$\mu_i,\gamma$
&
$\mu_i\geq0$ and
$\gamma=\mu_1+\mu_2+\mu_3\geq0$; Dunkl multiplicities and their radial
combination.
\\
\hline
$\kappa,\alpha$
&
$\kappa>0$ is the inverse curvature radius; the Deng--Fan inverse-range
parameter is restricted to $\alpha=2\kappa>0$ in the exactly solvable
subfamily.
\\
\hline
$D_e,r_e,m_{\mathrm{red}}$
&
$D_e>0$, $r_e>0$, and $m_{\mathrm{red}}>0$; dissociation energy,
equilibrium separation, and reduced mass.
\\
\hline
$b$
&
$b=e^{2\kappa r_e}-1>0$.
\\
\hline
$\ell,\Lambda_{\ell,\gamma}$
&
$\ell\in\mathbb{N}_0$ and
$\Lambda_{\ell,\gamma}=(\ell+\gamma)(\ell+\gamma+1)\geq0$.
\\
\hline
$g_V,g_{\ell,\gamma}$
&
$g_V$ is the dimensionless inverse-square coefficient of the central
potential, and
$g_{\ell,\gamma}=\Lambda_{\ell,\gamma}+g_V$ is the complete local
inverse-square coupling.
\\
\hline
$\mathcal{A}_{\ell,\gamma}$
&
$\mathcal{A}_{\ell,\gamma}
=\hbar^2\kappa^2\Lambda_{\ell,\gamma}/(2m_{\mathrm{red}})
+D_e b^2/4>0$; coefficient of
$\operatorname{csch}^2(\kappa r)$.
\\
\hline
$\mathcal{B}_{\kappa}$
&
$\mathcal{B}_{\kappa}=D_e b(b+2)/2>0$; positive magnitude of the
$-\coth(\kappa r)$ term.
\\
\hline
$\mathcal{E}_{\mathrm{c}}$
&
$\mathcal{E}_{\mathrm{c}}
=\hbar^2\kappa^2(\gamma+1)^2/(2m_{\mathrm{red}})
+D_e(1+b+b^2/2)$; constant energy shift.
\\
\hline
$a_{\ell,\gamma}$
&
$a_{\ell,\gamma}
=2m_{\mathrm{red}}\mathcal{A}_{\ell,\gamma}/(\hbar^2\kappa^2)>0$;
complete dimensionless inverse-square coupling.
\\
\hline
$q_\kappa$
&
$q_\kappa
=2m_{\mathrm{red}}\mathcal{B}_{\kappa}/(\hbar^2\kappa^2)>0$;
dimensionless Eckart attraction.
\\
\hline
$E_{\mathrm{th}}$
&
$E_{\mathrm{th}}=\mathcal{E}_{\mathrm{c}}-\mathcal{B}_{\kappa}
=D_e+\hbar^2\kappa^2(\gamma+1)^2/(2m_{\mathrm{red}})$; continuum
threshold.
\\
\hline\hline
\end{tabular}
\endgroup
\end{table*}

\begin{acknowledgments}
The author acknowledges the University of Santo Tomas for its
institutional support and for providing an environment conducive to
this research.
\end{acknowledgments}

\section*{Data availability statement}

No external datasets were used. The numerical results can be
reproduced from the finite-difference matrix in
Eq.~\eqref{eq:numerical_tridiagonal_matrix}, the benchmark parameters
reported in Section~\ref{sec:numerical_verification}, and the grid
sizes given in Tables~\ref{tab:numerical_spectrum_convergence} and
\ref{tab:numerical_nonzero_dunkl_benchmark}.

\bibliography{ref}

\end{document}